\documentclass[aps, prstab, 
              reprint, 
              amsmath, amssymb, amsfonts, 
              groupedaddress, 
              twocolumn, 
              showpacs, 
              floatfix]{revtex4-1}

\usepackage{graphicx}
\usepackage{dcolumn}
\usepackage{bm}
\usepackage{amsmath,amssymb}
\usepackage{color}
\usepackage{verbatim}
\usepackage{float}
\usepackage[caption = false]{subfig}

\usepackage[breaklinks=true]{hyperref}
  \hypersetup{
  colorlinks=true,
  linkcolor=black,
  citecolor=black,
  filecolor=black,
  menucolor=black,
  urlcolor=black
  }
\usepackage{multirow}
\usepackage{tabularx}
\usepackage{array}

\begin{document}

\title{High-Energy and High-Quality Ion beams in Light Sail Acceleration}

\author{Maitreyi Sangal}
\affiliation{Max-Planck-Institut f\"ur Kernphysik, Saupfercheckweg 1, D-69117 Heidelberg, Germany}
\author{Matteo Tamburini}\email{matteo.tamburini@mpi-hd.mpg.de}
\affiliation{Max-Planck-Institut f\"ur Kernphysik, Saupfercheckweg 1, D-69117 Heidelberg, Germany}

\date{\today}

\begin{abstract}
A superintense laser pulse illuminating a thin solid-density foil can, in principle, accelerate the entire foil, therefore yielding dense, collimated, and quasi-monoenergetic ion beams. These unique features render radiation pressure acceleration in the light sail regime a promising acceleration mechanism suited for applications where dense and high-flux ion beams are required. However, the onset of several instabilities typically results into foil deformation and heating, which cause premature termination of the radiation pressure acceleration stage and strong broadening of the ion spectrum. Here we show that (i) a relation between the attainable ion energy per nucleon and the development of instabilities exists, such that increasing the ion energy results into an increase of transverse modulation effects and, (ii) that the above relation can be weakened with proper matching of laser pulse-foil parameters, such that high-energy dense ion beams with high-quality spectral features can be produced.
\end{abstract}

\maketitle

\section{Introduction}

Radiation pressure acceleration (RPA) of a thin plasma foil with superintense laser pulses, also known as Light-Sail Acceleration (LSA), has attracted considerable interest over the last two decades owing to its envisaged  unique features~\cite{AMacchirev, HDaido}. In fact, the expected ultrashort beam duration and the relatively large number of accelerated particles may provide dense, high flux, high brilliance and low emittance ion beams. Moreover, the laser-to-beam energy conversion efficiency is predicted to steadily increase with increasing driver laser pulse energy and, in principle, approaches unity in the ultrarelativistic regime~\cite{AMacchiPRL09, AMacchiRPA}. The potential applications of such ion beams range from fast ignition~\cite{Tabak, JCFernandez} and plasma diagnostic techniques~\cite{BorghesiProbe, Borghesi_2008} to ion beam therapy~\cite{MBorghesiIBT, ULinzRev} as well as the production of radioactive materials~\cite{Fritzler, PMcKennaPRE}. 

Early experiments already demonstrated the generation of multi-MeV proton and ion beams showing fair agreement both with the predicted quasi-monochromatic features and with the ion energy scaling law as a function of the laser and plasma parameters~\cite{AHenig, SKarMultsp, Aurand_2013, CAJPalmerRTIExp, SsteinkeStable, JHBin}. Currently, the advent of multipetawatt lasers in facilities such as ELI~\cite{EliURL}, Apollon~\cite{ApollonURL, apollon} and XCELS~\cite{xcelsURL} paves the way to soon explore the relativistic LSA regime experimentally. 
However, further studies also showed that detrimental effects such as the onset of instabilities~\cite{FPegoraroBubbles, ZXiaomei, Eliasson2015, YWanPhysMech}, and finite spot size effects~\cite{FDollar} lead to foil deformation and electron heating, which ultimately result into foil disruption with strong ion spectrum broadening and the premature termination of the RPA phase. 

For the above-mentioned reasons, a number of ingenious methods were put forward, e.g., to enhance the maximal achievable ion energy via phase locking~\cite{SVBulanov}, cocoon formation and pulse self-focusing~\cite{MTamburiniRR}, laser pulse profiling~\cite{SSBulanov} or guiding~\cite{BulanovGuid} which, however, also resulted into a decrease of the number of accelerated ions and a noticeable deterioration of the quasi-monochromatic features of the ion spectrum. Moreover, although smart techniques such as laser pulse shaping~\cite{BulanovFlattop, MChenLaser} or foil engineering~\cite{MChenShaped, MChenStabilised, Tyu, XFShen} reduce foil deformation and stabilize RPA acceleration while circularly polarized laser pulses quench both electron heating~\cite{AMacchiIonBunches, AMacchiRPA, APLRnjp, OKlimo, APLRppcfHB, LiseykinaCP} and energy dissipation due to radiation reaction effects~\cite{Tamburini_2010}, the predicted ion spectrum remains relatively broad. Thus, for many applications, marked improvements are required in two major directions simultaneously, namely (i) increase the ion energy per nucleon up to hundreds MeV and, (ii) improve the monochromatic features of the ion spectrum.

Here we show that an intimate connection exists between the attainable LSA average energy per nucleon and the development of the transverse Rayleigh-Taylor-like instability. Thus, the generation of high-energy and also high spectral quality ion beams requires to mitigate the above-mentioned relation, which can be realized with proper choice of the laser and foil parameters. In particular, we demonstrate that substantial improvements of the quasi-monochromatic features of the ion spectrum are possible either by employing a train of short and intense laser pulses or by a single short and intense laser pulse accelerating a double-layer foil. Indeed, proper matching of the laser pulse and foil parameters can yield ion beams with higher energy than that predicted by the light sail model and, simultaneously, a markedly improved ion beam spectrum.

\section{The Light Sail Model \& Transverse Instabilities}

In LSA, the plasma foil is modeled as a flat rigid `mirror' with mass density $\rho$ and thickness $\ell$ accelerated by the radiation pressure of a circularly polarized intense plane-wave-like laser pulse with frequency $\omega$. In the laboratory frame, the resulting light sail equation for a laser pulse normally incident on the foil is (see, e.g., Refs.~\cite{Esirkepov, AMacchiPRL09, AMacchiRPA})
\begin{equation} \label{ls_eq}
\frac{d}{dt}(\gamma \beta) = \frac{2I(t')}{\rho \ell c^2} R(\omega')\left(\frac{1 - \beta}{1 + \beta}\right),
\end{equation}
where $\beta=V/c$ is the foil velocity in units of the speed of light in vacuum $c$, $\gamma = (1 - \beta^2)^{-1/2}$ is the foil relativistic factor, $I(t')$ is the laser intensity at the foil position as a function of the retarded time $t'= t - X(t)/c$, where $X(t)$ is the instantaneous foil position, and $R(\omega')$ is reflectivity of the foil as function of the foil rest frame frequency $\omega' = D \omega$, with $D = \sqrt{(1-\beta)/(1+\beta)}$ being the Doppler factor. By changing from time $t$ to the retarded time $t'$, Eq.~(\ref{ls_eq}) can be solved exactly for arbitrary laser pulse temporal profile. For $R(\omega') \approx 1$, the predicted energy per nucleon $\varepsilon_u$ is~\cite{AMacchiRPA}
\begin{equation} \label{eu}
\varepsilon_u = \frac{\mathcal{E}^2}{2(1+\mathcal{E})} m_u c^2,
\end{equation}
where $m_u$ is the atomic mass unit and $\mathcal{E} = 2 \mathcal{F}/\rho \ell c^2$ is basically the ratio between the laser pulse fluence $\mathcal{F}=\int{I(t') dt'}$, i.e., the laser pulse energy per unit surface, and the foil surface density $\rho \ell$. 

From Eq.~(\ref{eu}) it immediately follows that, in principle, arbitrarily high ion energy is attainable by increasing the laser fluence or reducing the foil surface density, provided that $R(\omega') \approx 1$ during acceleration. This last condition can be easily assessed by employing normalized units, i.e., by rewriting $\mathcal{E}$ as a function of the normalized laser field amplitude $a(\phi) = \sqrt{I(\phi)/I^*}$ and the normalized surface density $\zeta = \pi n_e \ell / n_c \lambda$, where $I^* = m_e^2 \omega^2 c^3 / 4 \pi e^2$, $\phi = [t/T - X(t)/\lambda]$ is the laser phase at the foil position, and $n_c = m_e \omega^2 / 4 \pi e^2$ is the critical plasma density. Here $n_e$ is the electron number density, $e$ and $m_e$ are the electron charge and mass, while $T=2\pi/\omega$ and $\lambda = c T$ are the laser period and wavelength, respectively. For simplicity, we consider a foil with a single fully ionized atomic species, then
\begin{equation} \label{E}
\mathcal{E} = \frac{2 \pi Z m_e \int{a^2(\phi)\,d\phi}}{A m_u \zeta} = \frac{3 \pi Z m_e a^2_0 \tau}{4 A m_u \zeta}
\end{equation}
where $Z$ ($A$) is the atomic number (mass), and $a_0$ is the peak value of $a(\phi)$. In the last equality of Eq.~(\ref{E}) we considered, for definiteness, a $\sin^2$ laser pulse field (i.e., $\sin^4$ intensity) temporal profile with $\tau$ total duration in units of the laser period. Now, the foil reflectivity $R(\omega')$ is approximately unity when $a(\phi) < \zeta'(\phi)$~\footnote{Note that in the ultrarelativistic case the relation $a(\phi) < \zeta'(\phi)$ allows to increase the laser intensity or decrease the foil areal density maintaining $\mathcal{R} \approx 1$, which, in principle, allows to attain ion energies well beyond those attainable with the bound $a_0 \lesssim \zeta$.}, where $\zeta'(\phi) = \zeta/D(\phi)$ is the Doppler increased normalized foil surface density, which is automatically satisfied when $a_0 \lesssim \zeta$ (see Ref.~\cite{AMacchiRPA}). Thus, from Eq.~(\ref{E}) $\mathcal{E} \lesssim 3 \pi Z m_e a_0 \tau/4 A m_u$ and, for given laser pulse fluence $\mathcal{F} \propto a_0^2 \tau$, higher ion energies are attainable by decreasing the laser pulse intensity $I \propto a_0^2$ and, correspondingly, by increasing the laser pulse duration $\tau$. 

In addition, a lower laser intensity with longer laser pulse duration also results into (i) a more adiabatic foil acceleration, which maintains the plasma foil compact during the acceleration phase and, (ii) a reduced foil heating and, consequently, a suppressed expansion of the foil driven by energetic electrons. In fact, within the thin foil approximation~\footnote{The key condition for this approximation to hold is $n_e \gg a_0 n_c$ and $\ell \ll \lambda$.}, the `transverse' relativistic factor $\Gamma$ associated to the transverse relativistic motion of the electrons of the foil is~\cite{AMacchiRPA}
\begin{align}
\Gamma^2(\phi) = & \big\lbrace 1 + a^2(\phi) - \zeta'^2(\phi) + \big[ (1 + a^2(\phi) - \zeta'^2(\phi))^2 \nonumber \\
+ &  4 \zeta'^2(\phi) \big]^{1/2} \big\rbrace / 2
\end{align}
such that $\Gamma \approx 1$ ($\Gamma \approx \sqrt{a_0}$) when $a_0 \ll \zeta$ ($a_0 \approx \zeta \gg 1$) due to the suppressed (increased) foil transmissivity. As it will be clear below, 1D particle-in-cell (PIC) simulations confirm the two above-mentioned conclusions. In particular, for fixed foil parameters and given laser fluence, 1D PIC simulations show improved monochromatic features of the ion spectrum when longer and less intense laser pulses are employed [see Fig.~\ref{fig:1}(a)]. 

However, the obtained 1D results drastically change in a multi-dimensional geometry, even when plane-wave laser pulses and flat foils are employed, such that finite size effects are absent. In fact, the accelerated foil is subject to a number of instabilities such as Weibel-like instability~\cite{ZXiaomei}, two-stream-like instability~\cite{YWanPhysMech}, and Rayleigh-Taylor-like (RT) instability~\cite{FPegoraroBubbles, CAJPalmerRTIExp}. Indeed, our high-resolution 2D PIC simulations show that transverse density modulations with a broad spectrum of wavelengths $\lambda_m$ are present, whose relative importance also depends on the temporal intensity profile of the laser pulse (see below). However, relatively small perturbations $\lambda_m \ll \lambda$ tend to be `smoothed' in the laser pulse-foil interaction, whereas longer wavelength modes $\lambda_m \gtrsim \lambda$ are amplified~\cite{CAJPalmerRTIExp}. The linear theory of RT instability predicts that the shorter wavelength mode compatible with diffraction effects should dominate~\cite{FPegoraroBubbles}, which is indeed confirmed by experiments where $\lambda_m \sim \lambda$ was shown to be the dominant wavelength of density modulations~\cite{CAJPalmerRTIExp}. Thus, the dominant $\lambda_m \sim \lambda$ mode of RT instability is a key quantity for assessing the importance of transverse effects.

In its linear phase, RT instability grows exponentially, with its exponent rising linearly with time in the nonrelativistic regime. By contrast, the RT exponent grows proportionally to the cubic root of time in the ultrarelativistic regime~\cite{FPegoraroBubbles}. The drastically different scaling in the nonrelativistic and ultrarelativistic regime suggested that a rapid acceleration of the foil to ultrarelativistic velocity could effectively stop the growth of the instability~\cite{FPegoraroBubbles}. However, in the ultrarelativistic regime the duration of the laser-foil interaction is also considerably longer, because the laser pulse and the foil basically move with the same velocity, such that the interaction lasts for a time considerably longer than the laser pulse duration. Thus, the net effect of the growth of the RT instability is more appropriately analyzed as a function of the laser phase at the foil position $\phi$ instead of time $t$. 
For the above-considered fully ionized single species foil accelerated by a $\sin^2$ laser field temporal profile, the RT instability grows as $\exp[\Phi]$ with~\cite{FPegoraroBubbles}
\begin{equation} \label{grow}
\Phi = 2 \pi \int{d\phi \left[\frac{Z m_e a^2(\phi)\lambda}{A m_u \zeta \lambda_m}\right]^{1/2}} = \frac{\pi a_0 \tau}{\sqrt{\zeta}} \sqrt{\frac{Z m_e \lambda}{A m_u \lambda_m}},
\end{equation}
which implies $\Phi \lesssim \pi \sqrt{\zeta} \tau \sqrt{Z m_e/ A m_u}$ for $a_0 \lesssim \zeta$ and $\lambda_m \sim \lambda$. By comparing Eqs.~(\ref{eu})-(\ref{E}) with Eq.~(\ref{grow}), it immediately follows that (i) an increase of ion energy per nucleon is necessarily accompanied by an increase of the growth of the RT instability, (ii) attaining larger ion energy with minimal RT instability development favors a larger $Z/A$ ratio which is attained, e.g., by employing hydrogen instead of carbon, and (iii) for fixed laser fluence and foil parameters, a simultaneous increase of the laser intensity and decrease of the laser pulse duration mitigates RT instability.

Note that the above dependence on the laser pulse intensity and duration is opposite to what has been obtained above for the 1D dynamics, i.e., when no transverse effect exists. This opposite trend is confirmed by comparing 1D and 2D PIC simulations with the same parameters [see Figs.~\ref{fig:1}(a)-\ref{fig:1}(b)]. On one hand, these results suggest that an optimal region of laser pulse intensity and duration exists, where the foil acceleration is sufficiently adiabatic and transverse instabilities are sufficiently mitigated to maintain uniform RPA of the foil. On the other hand, it suggests that one needs to relax the relation between $\mathcal{E}$ and $\Phi$ in order to simultaneously increase the final ion energy while preserving the monochromatic spectral features of ideal LSA. This can be done by going beyond the simple LS model and by allowing more complex laser pulse and foil parameters.

\section{PIC Simulation results}

\subsection{Laser Pulse Duration-Intensity Relation}

In order to test the above predictions on the relation between the laser pulse intensity and the laser pulse duration, we carried out 1D and 2D PIC simulations with the fully relativistic and fully parallel PIC code Smilei~\cite{SmileiPaper}. In all our simulations, the laser pulse is circularly polarized with $\sin^2$ temporal field envelope and $\lambda=0.8\,\mu$m wavelength. The considered duration of laser pulses is $\Delta_\tau =$ 5, 10, 20, 30~fs full width at half maximum (FWHM) of the pulse intensity with normalized laser amplitude $a_0 \approx$ 71, 50, 35, 29, respectively. The above-mentioned laser pulse duration and field amplitude are chosen such that the laser pulse fluence is the same in all considered cases. The foil is initially composed of neutral carbon with thickness $\ell = 0.056\lambda$. The foil is fully ionized by the laser pulse field at the beginning of the interaction, where its electron density reaches the maximum value of 400~$n_c$. Note that the foil thickness is chosen to satisfy the optimal LSA condition $a_0 \approx \zeta$ for $a_0 \approx 71$, such that $a_0 \lesssim \zeta$ and the reflectivity is approximately unity in all considered cases. 

In our 2D simulations the foil is initially flat, and the laser pulse is modeled as a plane wave. Thus, no finite size effect is present, and a direct comparison with 1D results is possible. To accurately resolve the plasma dynamics, in 2D the simulation box is $16\lambda (x) \times 16\lambda (y)$ with $16000 (x) \times 16000 (y)$ grid points, and 66 (400) particles-per-cell are used for ions (electrons when full ionization is reached). The same spatial resolution and particles-per-cell are used in the corresponding 1D simulations.

\begin{figure}[t]
\includegraphics[width = 0.48\textwidth]{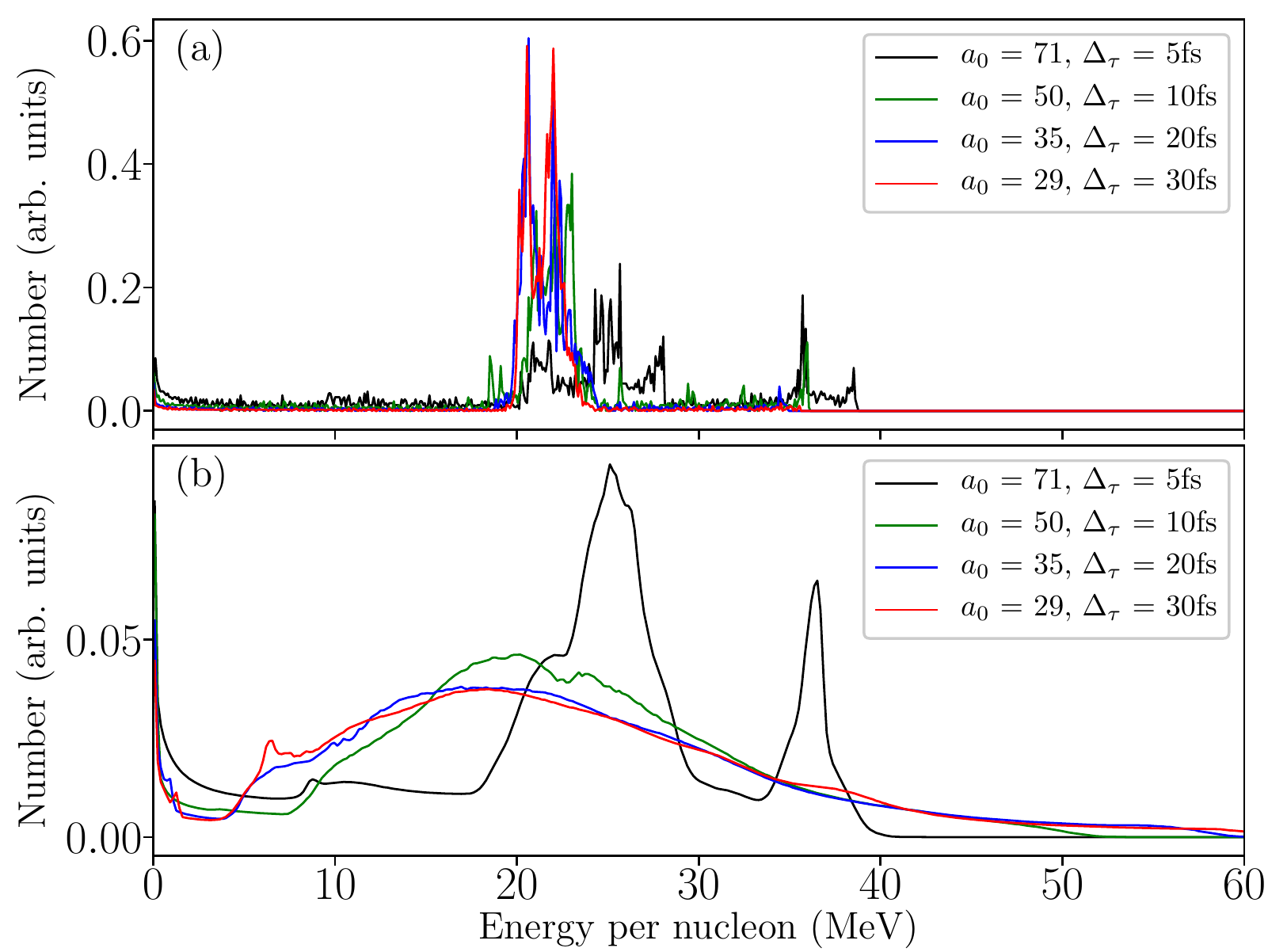}
\caption{\label{fig:1} Ion energy spectrum per nucleon for an initially neutral carbon foil with thickness $\ell = 0.056\lambda$ accelerated by a superintense laser pulse with duration $\Delta_\tau =$ 5, 10, 20, 30~fs and normalized amplitude $a_0 \approx$ 71, 50, 35, 29, respectively. (a) 1D PIC results, (b) 2D PIC results.}
\end{figure}

Figure~\ref{fig:1}(a) and Fig.~\ref{fig:1}(b) report the results obtained with 1D and 2D simulations, respectively. Figure~\ref{fig:1}(a) shows that the ion spectrum qualitatively improves with increasing laser pulse duration, with the ion energy per nucleon being in good agreement with the LS model prediction $\varepsilon_u \approx 21$~MeV. This occurs because the acceleration process is increasingly more adiabatic for longer duration and lower intensity. In fact, during the acceleration process the equilibrium between electrostatic force and radiation pressure leads to the cyclic acceleration of ion populations at the front surface of the foil, which results into the formation of loops in the ion phase space~\cite{AMacchiIonBunches, MGrechRPA}. For long and less intense pulses the momentum difference of these populations is small with the foil remaining compact. By contrast, for short and intense pulses the populations at the foil front are violently pushed forward and may even overshoot the foil consequently forming distinct separated energy populations as show in Fig.~\ref{fig:1}(a) for the 5~fs case. In addition, the larger $a_0$ for shorter pulses increases the laser pulse penetration into the foil resulting into increased electron heating, and also into a stronger effect of Coulomb explosion associated to the larger electron-ion spatial separation during acceleration.

Although the above 1D effects are still present in 2D simulations, in this case the foil is also subject to instabilities that result in transverse density modulations. Notably, these density modulations depend on the temporal envelope of the laser pulse (see below). In spite of the initial planar symmetry of the laser pulse and of the foil, Fig.~\ref{fig:1}(b) shows that 2D spectra are noticeably broader than the corresponding 1D spectra. In addition, in consistence with the expectation of stronger RT instability for longer duration pulses, a moderate improvement of the ion spectrum quality with decreasing laser pulse duration is visible for all cases. Remarkably, for the 5~fs duration case the ion spectrum is markedly improved and its features are much closer to those of the corresponding 1D simulation [see Fig.~\ref{fig:1}(b)]. This can be explained by noting that in almost all cases but the shorter duration pulse, $\Phi$ is well above unity ($\Phi>3$), which implies that RT is saturated and fully in the nonlinear regime (see Tab.~\ref{tab:1}). Thus, strong transverse foil modulations with dense clumps separated by lower density regions have already formed, with the laser pulse penetrating and heating the electrons of the lower density regions [see Fig.~\ref{fig:3}(a)]. In this case, no compact foil is still present, such that the exact value of $\Phi$ is no longer decisive for assessing the quality of the ion spectrum.

The above results on the relation between the laser pulse intensity and its duration imply that attaining relativistic energy per nucleon and high-quality ion beams in LSA requires smart techniques to suppress transverse instabilities and also non-adiabatic acceleration effects, simultaneously. In the following we consider two possible strategies, one based on laser pulse modulation, and the other one on optimal laser pulse-foil parameter matching.
\begin{table}[b]
\centering
\begin{tabular}{|c|c|c|c|c|c|}
\hline
$a_0$ & $\Delta_\tau$ (fs) & $\Phi$ & $\varepsilon_p$ (MeV) & $\Delta\varepsilon_p$ (MeV) & $\Delta\varepsilon_p/\varepsilon_p$ \\
\hline
$71$ & $5$ & $2.3$ & $25.2$ & $4.4$ & $0.17$ \\
\hline
$50$ & $10$ & $3.2$ & $21.0$ & $18.6$ & $0.89$ \\
\hline
\multirow{2}{*}{$35$} & $20$ & $4.5$ & $20.0$ & $23.2$ & $1.20$ \\
\cline{2-6}& $2 \times 10$ & $2.3 (1^{st})$ & $15.3$ & $2.6$ & $0.17$ \\
\hline
\multirow{2}{*}{$29$} & $30$ & $5.5$ & $19.0$ & $26.0$ & $1.37$  \\ 
\cline{2-6}& $3 \times 10$ & $1.8 (1^{st})$ & $20.4$ & $4.1$ & $0.20$ \\
\hline
\end{tabular}
\caption{\label{tab:1} Average ion energy per nucleon $\varepsilon_p$ within one FWHM energy range around the peak, FWHM of the ion peak $\Delta\varepsilon_p$ and relative energy spread $\Delta\varepsilon_p/\varepsilon_p$ from 2D PIC simulations of the interaction of a plane-wave laser pulse (train of 2 or 3 laser pulses) normalized amplitude $a_0$ and duration $\Delta_\tau$ with a flat carbon foil with thickness $\ell$ = $0.056\lambda$. The LS prediction of the ion energy per nucleon is 21~MeV, while the exponent of the RT instability growth $\Phi$ is given by Eq.~\ref{grow}.}
\end{table}
\begin{figure}[th]
\includegraphics[width = 0.48\textwidth]{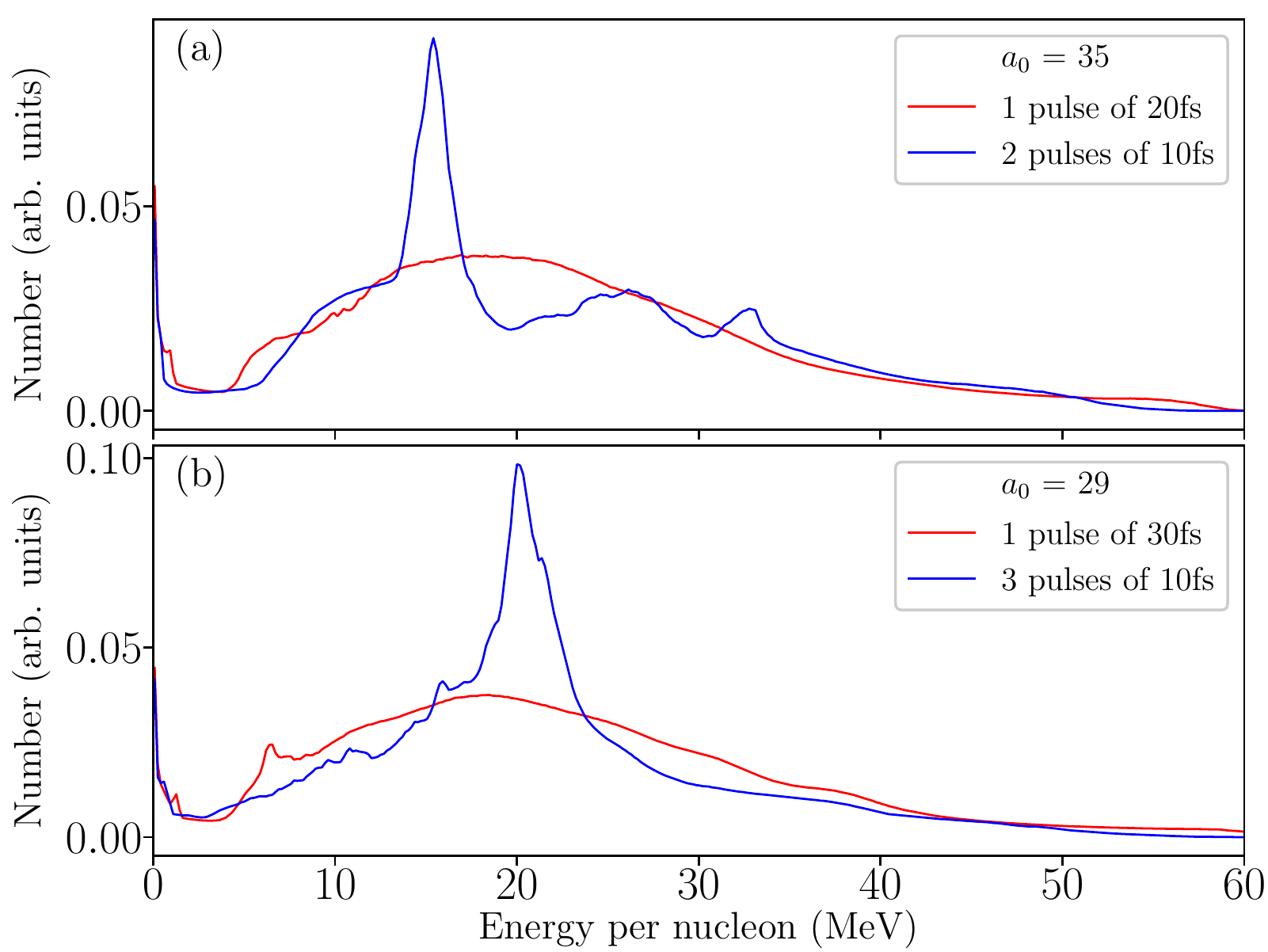}
\caption{\label{fig:2} Ion energy spectrum per nucleon for an initially neutral carbon foil with thickness $\ell = 0.056\lambda$ accelerated by (a) a single 20~fs (red line) and two 10~fs (blue line) laser pulses all with $a_0 \approx 35$, (b) a single 30~fs (red line) and three 10~fs (blue line) laser pulses all with $a_0 \approx 29$.}
\end{figure}

\subsection{Train of short pulses}

The first strategy consists in accelerating the foil with a train of short and intense laser pulses where, ideally, each single laser pulse of the train is such that $\Phi\ll 1$ for the considered foil parameters and, also, longitudinal non-adiabatic effects are subdominant. The rationale is that, in this case, foil density modulations induced by each laser pulse during RPA are sufficiently small that the electrons and ions of the foil can reorganize to re-establish quasi-neutrality and diffuse from the higher density regions to the lower density regions. This may result into foil `smoothing', i.e., the density modulations induced during the acceleration phase are suppressed before the following laser pulse interacts with the foil. In practice, it might be difficult to obtain all the above conditions simultaneously. However, here we show that even when each laser pulse of the train is such that $\Phi\approx2$ a substantial improvement of the ion energy spectrum is attainable compared to the single pulse case. We mention that ion acceleration with a train of two short and intense laser pulses has been already experimentally realized in the context of  target normal sheath acceleration, where a significant spectral enhancement compared to the case of a single laser pulse with the same total energy was observed~\cite{KMarkey}. 
\onecolumngrid
\begin{center}
\begin{figure}[t]
\includegraphics[width = 0.98\textwidth]{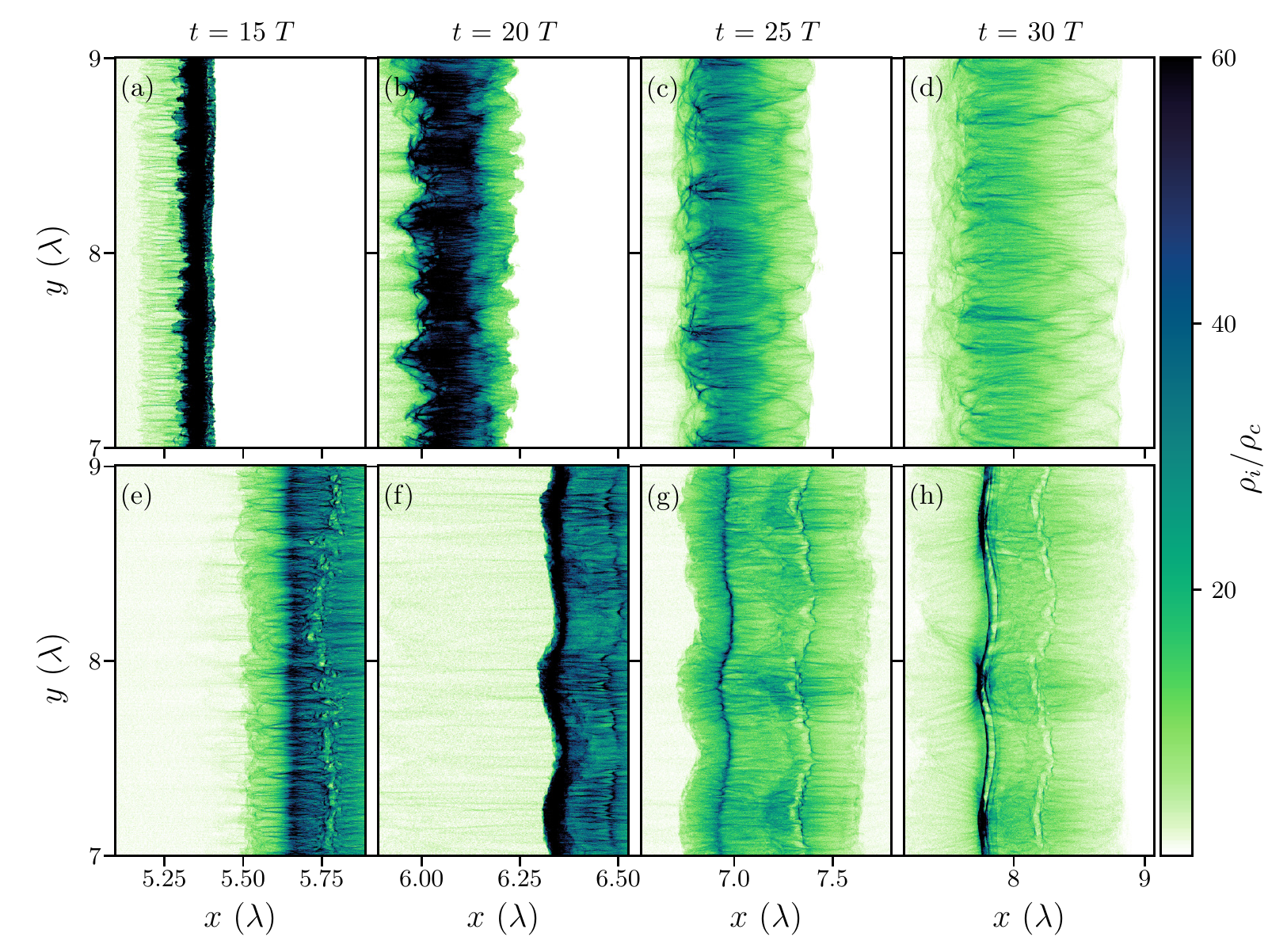}
\caption{\label{fig:3} Snapshots of the ion charge density $\rho_i$ during the laser pulse-foil interaction, which starts at $t=0$. Snapshots are taken at 15~$T$, 20~$T$, 25~$T$ and near the end of the laser pulse-foil interaction at 30~$T$. (a)-(d) single 30~fs pulse, (e)-(h) three 10~fs pulses.}
\end{figure}
\end{center}
\twocolumngrid
\onecolumngrid
\begin{center}
\begin{figure}[t]
\includegraphics[width = 0.98\textwidth]{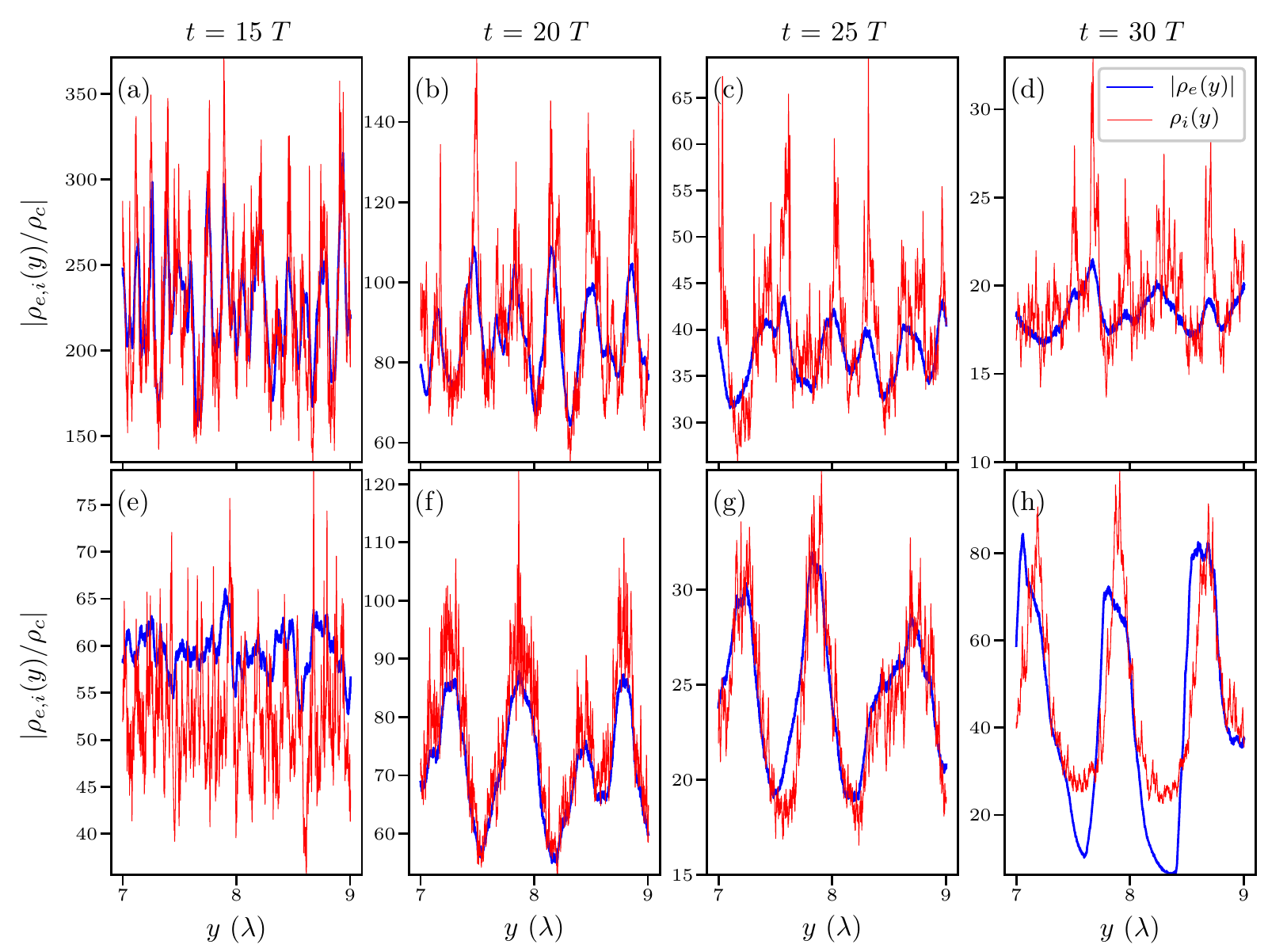}
\caption{\label{fig:4} Transverse electron $\vert\rho_{e}\vert$ (blue line) and ion $\rho_{i}$ (red line) charge density averaged over the higher density ion layer (see Fig.~\ref{fig:3}) in units of $\rho_c$ = $\vert e \vert n_c$. (a)-(d) single 30~fs pulse, (e)-(h) three 10~fs pulses.}
\end{figure}
\end{center}
\twocolumngrid

In the following we report the results of 2D simulations for a `train' of two and three identical 10~fs laser pulses and compare these results with those of a single 20~fs and 30~fs laser pulse, respectively. For each simulation, the $\sin^2$ shape, plane-wave transverse profile, total fluence and total duration as well as the peak intensity, or equivalently $a_0$, is the same for the train of pulses and for the single pulse. The foil and numerical parameters are the same as in previous simulations.

Figure~\ref{fig:2}(a) [Figure~\ref{fig:2}(b)] shows the ion spectra obtained with a single 20~fs (30~fs) pulse and with a train of two (three) 10~fs laser pulses.  The average ion energy per nucleon is in all cases in agreement with the LS model prediction, but the quality of the spectrum is markedly improved in the case of a train of laser pulses. In particular, a sharp peak is present in the case of multiple pulses. The relative spectral width $\Delta\varepsilon_p/\varepsilon_p \approx 0.17$ (0.20) of the peak in the spectrum for the case of two (three) pulses of 10~fs is significantly smaller compared to $\Delta\varepsilon_p/\varepsilon_p \approx 1.20$ (1.37) for a single 20~fs (30~fs) laser pulse, as summarized in Table~\ref{tab:1}. The fraction of accelerated ions is also larger for the case of a pulse train, where less ions are present at lower energies (see Fig.~\ref{fig:2}). Here $\Delta\varepsilon_p$ is the FWHM of the peak in the ion spectrum and $\varepsilon_p$ is the average ion energy per nucleon obtained averaging over $\pm\Delta\varepsilon_p/2$ around the peak.

\begin{center}
\begin{figure}[t]
\includegraphics[width = 0.98\textwidth]{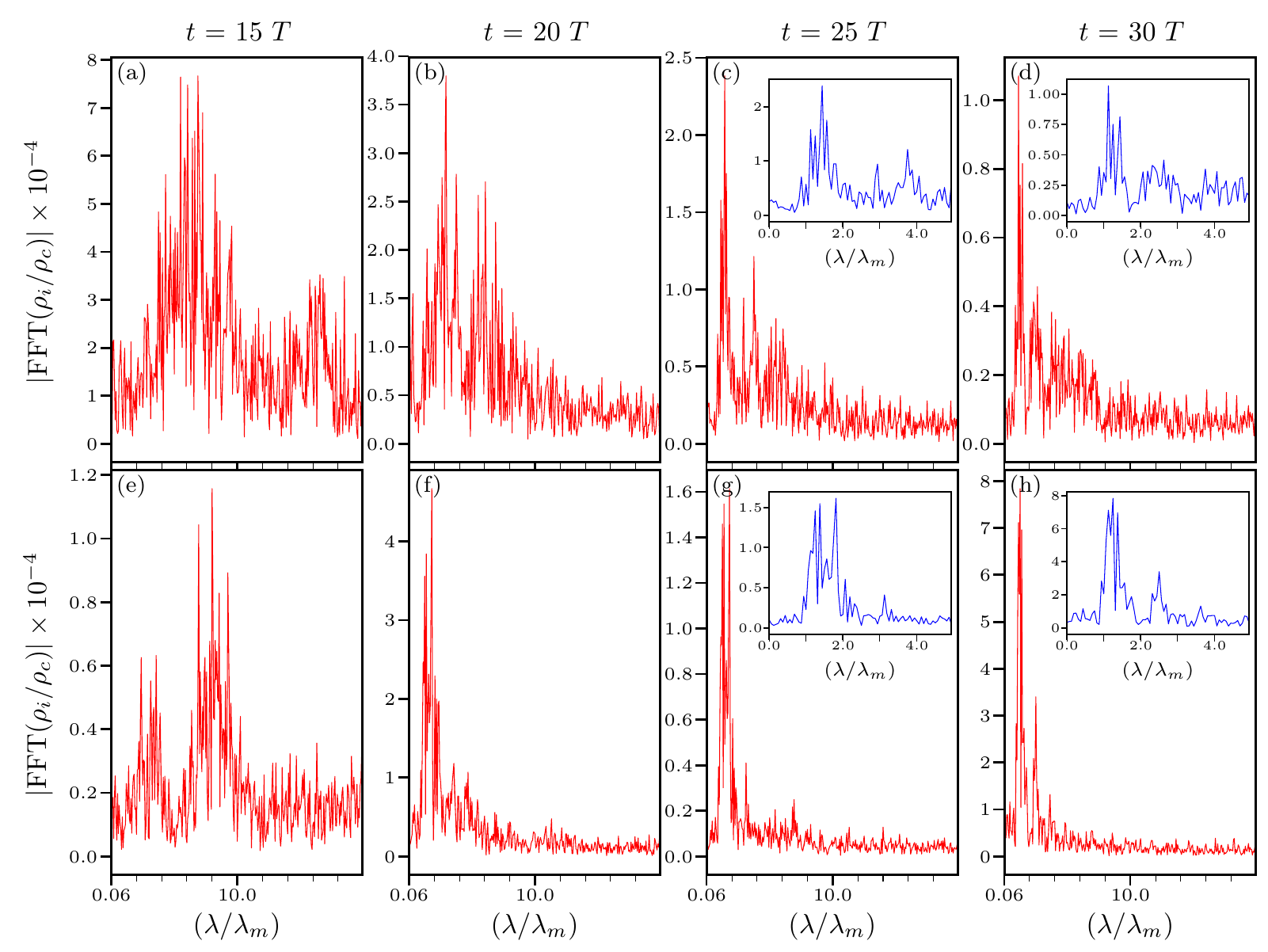}
\caption{\label{fig:5} Fourier transform of the normalized transverse ion charge density $\vert \text{FFT}(\rho_{i}/\rho_{c})\vert$ averaged over the higher density ion layer taken at 15~$T$, 20~$T$, 25~$T$ and 30~$T$ (see Fig.~\ref{fig:4}). (a)-(d) single 30~fs pulse, (e)-(f)] a train of three 10~fs pulses. The insets in panels (c)-(d) and (g)-(h) (blue lines) report a zoom of the longer wavelength mode $\lambda_m$ region, where $\lambda$ is the laser wavelength.}
\end{figure}
\end{center}

More insights can be gained by considering the ion and electron dynamics during acceleration. Figure~\ref{fig:3} displays snapshots of the carbon ion charge density both for a single 30~fs pulse [Figs.~\ref{fig:3}(a)-\ref{fig:3}(d)] and a train of three 10~fs pulses [Figs.~\ref{fig:3}(e)-\ref{fig:3}(h)] at intervals of 5~$T$ starting from 15~$T$ after the laser-foil interaction begins. Note that the peak intensity of the single 30~fs pulse reaches the foil at approximately 15~$T$, while for the train of pulses this occurs three times at approximately 5~$T$, 15~$T$, and 25~$T$. Figure~\ref{fig:3} shows that, both for the single and for the train of pulses, at 15~$T$ small scale filamentary structures followed by a denser modulated layer are present at the front of the foil. The presence of small scale modulations is further confirmed by considering the mean transverse electron $\vert\rho_{e}(y)\vert$ and ion $\rho_{i}(y)$ charge densities, which are obtained by averaging the charge distribution over a length corresponding to the thickness of the ion layer displayed in Fig.~\ref{fig:3}. Figure~\ref{fig:4} displays $\vert\rho_{e}(y)\vert$ (blue line) and $\rho_{i}(y)$ (red line), while Fig.~\ref{fig:5} reports the modulus of the Fourier transform of $\rho_{i}(y)$.

The formation of structures with scale $\lambda_m$ much smaller than the laser wavelength $\lambda$ indicates that kinetic instabilities dominate during the initial stage of acceleration~\cite{ZXiaomei, YWanPhysMech}. However, small scale modulations are later suppressed and smoothed as the laser is insensitive to structures much smaller than its wavelength due to diffraction. In fact, at later stages ($t>25T$), modulations with scale comparable with $\lambda$ clearly dominate (see Fig.~\ref{fig:5}) as expected from RT instability~\cite{FPegoraroBubbles, CAJPalmerRTIExp}. Note that the suppression of small scale structures with the dominance of $\lambda_m \sim \lambda$ modes occurs earlier in the case of a train of pulses than in the case of a single pulse and is also much more pronounced  (see Figs.~\ref{fig:3}-\ref{fig:5} for $t\geq 20 T$). In fact, each of the pulses of the train accelerates the foil therefore creating an electron-ion charge separation and triggering instabilities. However, the induced foil modulations are relatively small for each pulse of the train compared to the single pulse case and, as the interaction with each pulse of the train finishes, the electrons and ions of the foil diffuse longitudinally and transversely to restore local charge neutrality. This diffusion process tends to suppress most of the previously generated small scale structures, while long scale modulations are less affected. Thus, the following pulse of the train interacts with a relatively homogeneous and quasi-neutral foil, which is effectively further accelerated by radiation pressure. This sequential acceleration process is effective provided that no long scale pre-plasma has formed before each pulse interacts with the foil, such that the foil is sufficiently compact and its reflectivity remains nearly unity.

\onecolumngrid
\twocolumngrid

\begin{figure}[t]
\centering
\includegraphics[width = 0.48\textwidth]{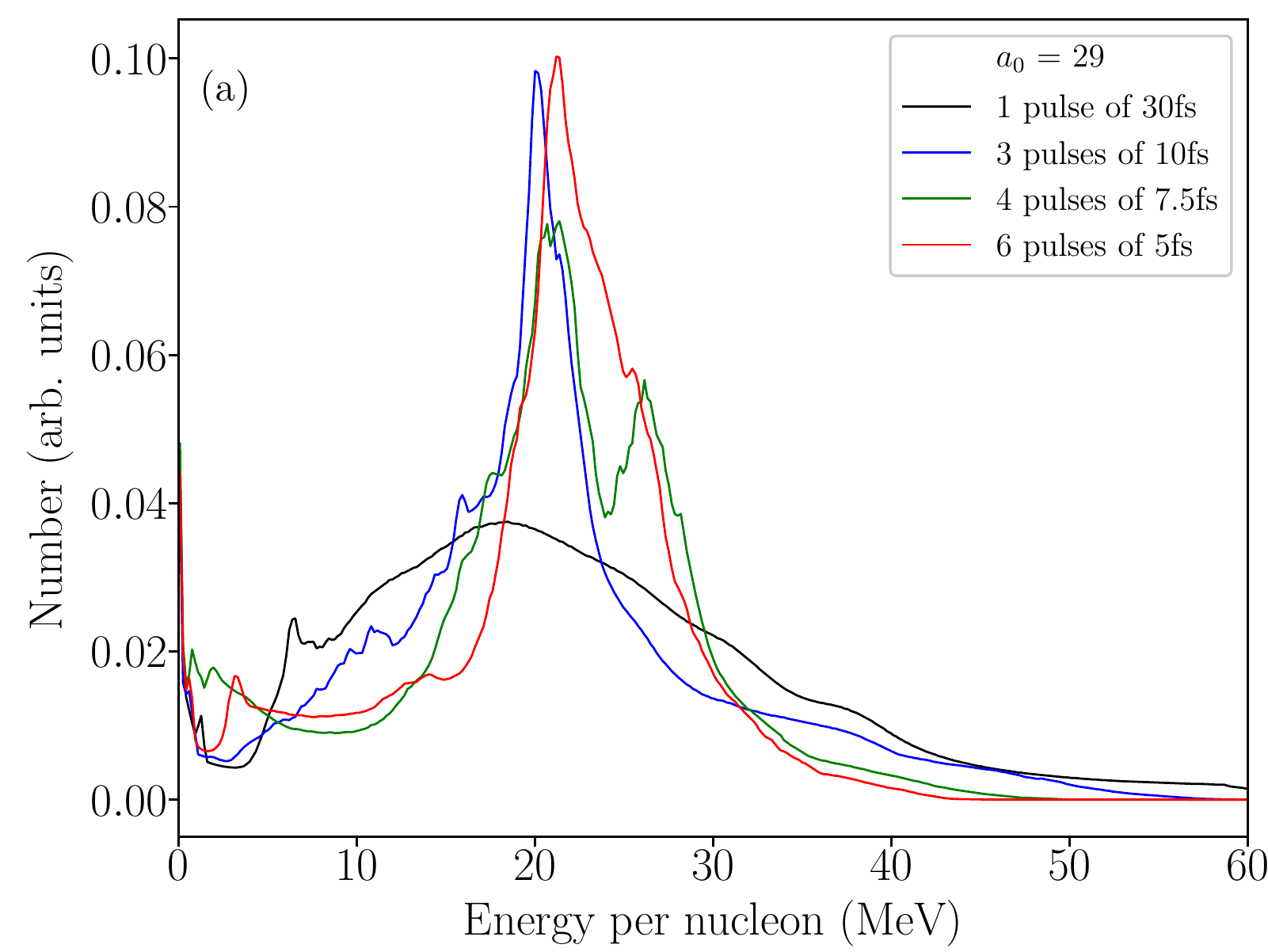}
\caption{\label{fig:6} Ion energy spectrum per nucleon for an initially neutral carbon foil with thickness $\ell = 0.056\lambda$ accelerated by a single 30~fs (black line) laser pulse, a train of three 10~fs (blue line), four 7.5~fs (green line), and six 5~fs (red line) laser pulses. The normalized amplitude of each pulse is $a_0 \approx 29$.}
\end{figure}

Finally, we consider the effect of increasing the number of laser pulses of the train with fixed total fluence and total duration. In this case each laser pulse of the train has the same peak amplitude $a_0$ as in previous simulations but smaller duration, such that $\Phi$ of each pulse decreases and transverse instabilities should be relatively suppressed. However, as discussed above, a sharp rise of the laser pulse intensity renders the acceleration process increasingly less adiabatic with the possible formation of multiple ion populations with different energies and increased electron heating. 
Figure~\ref{fig:6} shows the ion energy spectrum per nucleon for a single 30~fs pulse (black line) and a train of three 10~fs pulses (blue line), four 7.5~fs pulses (green line), and six 5~fs pulses (red line) all with the same total fluence, peak intensity and total duration. Fig.~\ref{fig:6} shows that increasing the number of laser pulses of the train reduces the number of ions at low energies and increases the total number of ions in the peak. However, the width of the peak does not reduce, and for the four 7.5~fs pulses the formation of a double peak is also observed, which is a signature of non-adiabatic acceleration. These results further confirm that high-quality spectral features are obtained when both transverse effects and non-adiabatic processes leading to the formation of multiple ion populations are suppressed.

\subsubsection{Finite spot size effects}

So far we have investigated the influence of transverse effects, mainly focusing on the dominant mode of RT instability, and of strong longitudinal gradients associated with non-adiabatic ion acceleration in the interaction of plane-wave laser pulses with flat foils. Naturally, in practice the laser pulse focal radius is of the order of a few micrometers, such that it is important to ascertain that the above findings also hold when finite spot size effects are present. 

It is known that the finite size of the focal spot may result into foil deformation, and that this can be prevented either using more transversely uniform super-Gaussian profiles~\cite{MChenLaser} or transversely modulated foils~\cite{MChenShaped}. Here we show that by employing transversely fourth order super-Gaussian profiles, similar results as the plane-wave pulse can be obtained, such that our previous conclusions can be readily extended also to realistic finite spot size laser pulses. In our simulations, the laser has $8\,\lambda$ diameter FWHM of the intensity, while all the other laser, foil as well as numerical parameters are the same as in previous simulations but the computational box size, which was increased to $20\lambda(x)\times16\lambda(y)$.
\begin{figure}[t]
\centering
\includegraphics[width = 0.48\textwidth]{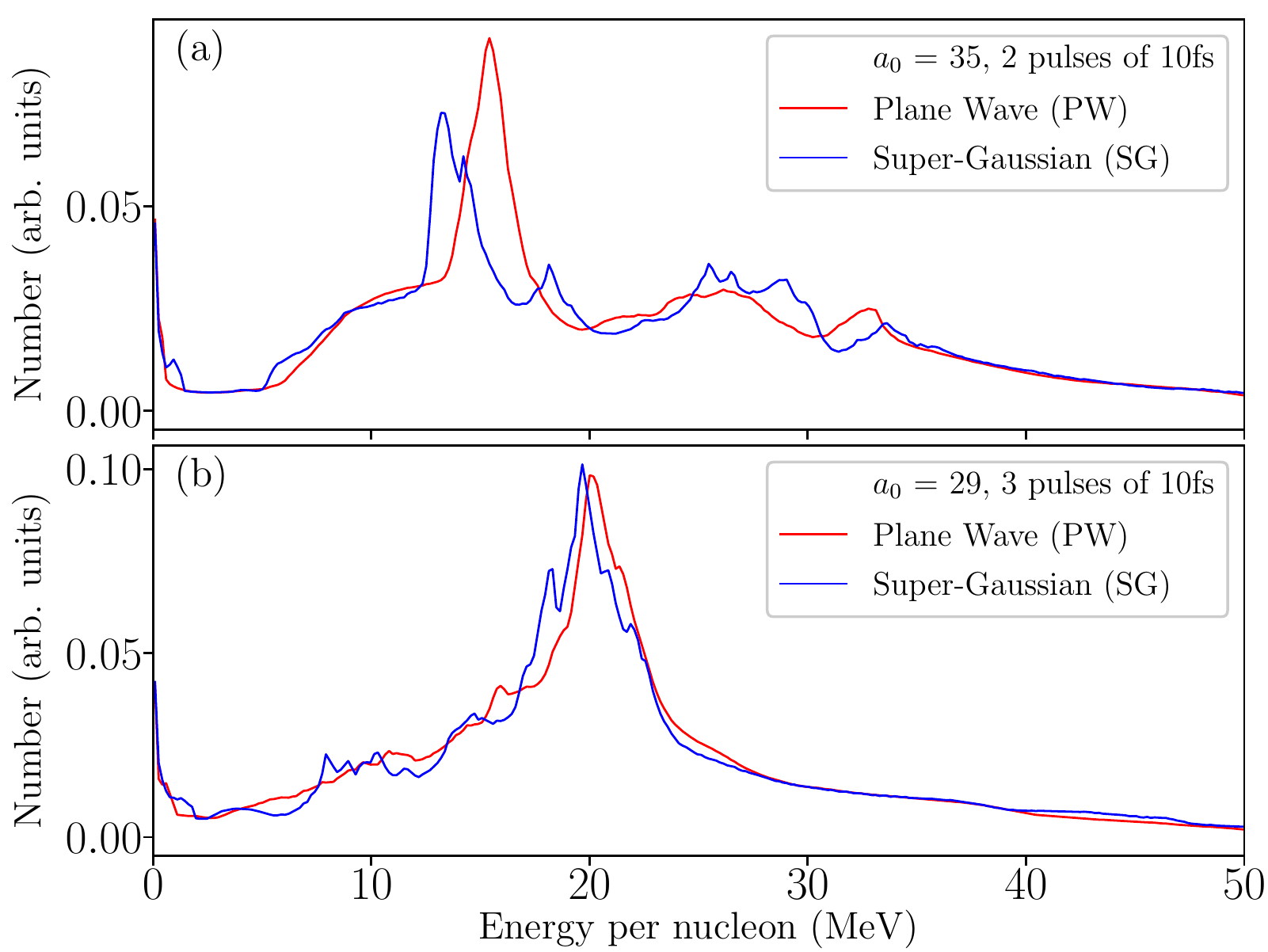}
\caption{\label{fig:7} Ion energy spectrum per nucleon for an initially neutral carbon foil with thickness $\ell = 0.056\lambda$ accelerated by a train of plane-wave (PW) laser pulses (red line) and by a train of super-Gaussian (SG) laser pulses (blue line) with $8\,\lambda$ diameter. For the supergaussian case the reported spectrum corresponds to ions within $2\,\lambda$ diameter around the laser axis. (a) two 10~fs laser pulses, (b) three 10~fs laser pulses.}
\end{figure}
Figure~\ref{fig:7}(a)-\ref{fig:7}(b) display the ion spectrum per nucleon for the train of two 10~fs pulses [Figure~\ref{fig:7}(a)] and three 10~fs pulses [Figure~\ref{fig:7}(b)] both for a plane-wave pulse (red line) and for a super-Gaussian pulse (blue line), where only ions within $2\,\lambda$ diameter around the laser axis are considered. A similar spectrum is observed both for the plane-wave and for the super-Gaussian laser pulses, therefore confirming that for sufficiently uniform transverse laser profiles, finite spot size effects are subdominant.

\subsection{Optimal laser pulse-foil parameter matching}

In the first part of our work we have shown that RPA with relatively long duration and low intensity laser pulses favors the suppression of longitudinal non-adiabatic effects. In sharp contrast, RPA with short duration and high intensity laser pulses favors the suppression of transverse effects. This opposite tendency has been confirmed by performing high-resolution 1D and 2D PIC simulations. In fact, for example, for the shorter considered duration $\Delta_\tau \approx 5\text{ fs}$ and higher intensity $a_0 \approx 71$ laser pulse, the obtained 1D and 2D spectra show nearly the same features [see the black line in Fig.~\ref{fig:1}(a) and Fig.~\ref{fig:1}(b)]. On one hand, the similarity between 1D and 2D spectra indicates that transverse effects are strongly suppressed. On the other hand, the presence of two distinct and well-separated peaks in the ion spectrum indicates that the sharp growth of the laser intensity resulted into a stronger acceleration of the front part of the foil than its rear part. In fact, the front region of the foil undergoes hole-boring acceleration with normalized velocity~\cite{PhysRevLett.102.025002}
\begin{equation} \label{vhb}
\beta_{\text{HB}}(t) = \frac{v(t)_{\text{HB}}}{c} = \frac{\sqrt{B(t)}}{1 + \sqrt{B(t)}}
\end{equation}
where $B(t) = I(t) /\rho c^3$. By contrast, at the beginning of the interaction $\beta_{\text{LS}}(t) \approx \mathcal{E}(t) = 2 \int_0^t{I(t') dt'} / \rho \ell c^2$,  where $\beta_{\text{LS}}(t)$ is the normalized light sail velocity. Thus, at the beginning of the interaction $\beta_{\text{HB}}(t) > \beta_{\text{LS}}(t)$ and the front part of the foil necessarily moves earlier and faster than the rear part of the foil until the condition $\beta_{\text{LS}}(t) \gtrsim \beta_{\text{HB}}(t)$ is reached. Thus, a sharp rise of the laser pulse intensity creates a first ion population originating from the foil front with significantly larger energy than that of the remaining part of the foil.
\begin{figure}[t]
\centering
\includegraphics[width = 0.48\textwidth]{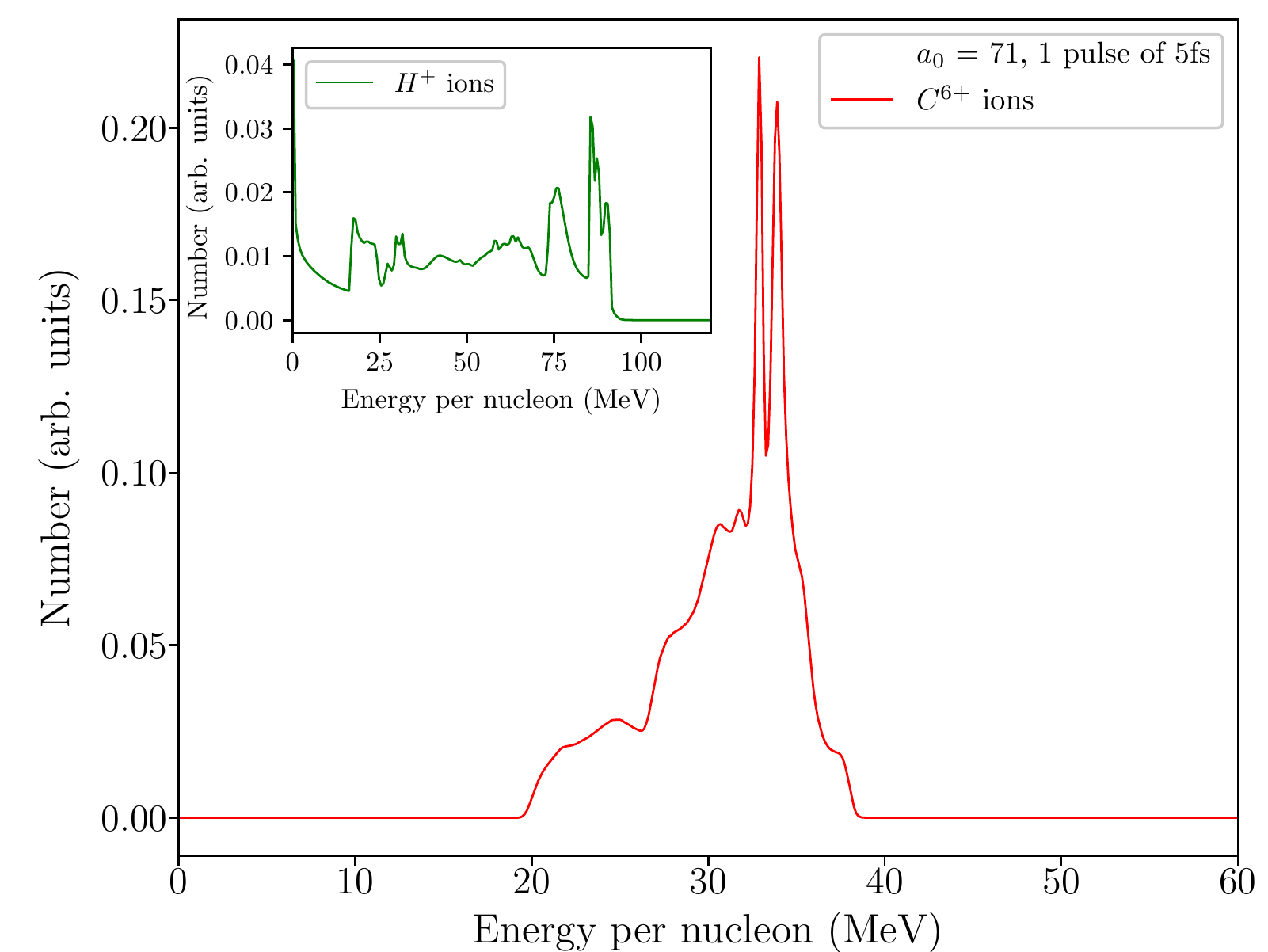}
\caption{\label{fig:8} Ion energy spectrum per nucleon for carbon (red line) and hydrogen (inset, green line) in the interaction of a single 5~fs laser pulse with normalized amplitude $a_0 \approx 71$ with a double-layer foil. The foil initially consists of a first layer of hydrogen with thickness $0.64\,\lambda$ and 20~$n_c$ electron density and a second layer of carbon with thickness $0.025\,\lambda$ and $400\,n_c$ electron density.}
\end{figure}

In order to attain high-energy and high-quality ion beams, it is therefore critical to retain the suppression of transverse effects while simultaneously reducing non-adiabatic effects that lead to the generation of distinct ion populations with noticeably different energies. Here we propose to employ short and intense laser pulses for RPA of a double-layer foil made of two distinct ion species with parameters chosen to both suppress transverse effects and achieve a smooth transition from the hole-boring to the light-sail stage. 
The laser pulse is the same as above, with a plane-wave transverse spatial profile, $\sin^2$ temporal field envelope, $a_0 \approx 71$ and $\Delta_\tau \approx 5\text{ fs}$. The numerical parameters are also the same as in the above-considered simulations. The foil has two layers, the first layer is made of initially neutral hydrogen with electron number density $n_{e_H} \approx 20\,n_c$ (when ionized) and $\ell_H = 0.64\,\lambda$ thickness, while the second layer is made of initially neutral carbon with electron number density $n_{e_C} \approx 400\,n_c$ (when fully ionized) and $\ell_C = 0.025\,\lambda$ thickness. For choosing the thickness of the first layer of the foil, the time $t_m$ at which $\beta_{\text{LS}}(t_m) \approx \beta_{\text{HB}}(t_m)$ was first determined assuming a hydrogen foil with normalized surface density $\zeta = a_0$ and solving Eq.~(\ref{ls_eq}) numerically. Then, the thickness of the foil was chosen to match the distance traveled by the hole-boring front in a time $t_m$, i.e., $\ell_H \approx c \int_0^{t_m}{\beta_{\text{HB}}(t) \,dt}$. Finally, the thickness of the second layer $\ell_C$ was chosen such that $\zeta = \pi (n_{e_H} \ell_H + n_{e_C} \ell_C) / n_c \lambda \approx a_0$. The above-mentioned conditions are such that the first layer undergoes non-adiabatic effects with the formation of multiple ion populations whereas the second layer is nearly uniformly accelerated via the LS mechanism since $\beta_{\text{HB}} \lesssim \beta_{\text{LS}}$ when the hole-boring front starts to interact with the second layer, also due to the fact that the second layer is denser than the first one.

Figure~\ref{fig:8} reports the obtained carbon (red line, main panel) and hydrogen (green line, inset) spectrum from our 2D PIC simulations. Whereas the hydrogen spectrum is very broad with multiple separated peaks corresponding to multiple ion populations, the carbon spectrum is much more narrow than in the case of a single species foil [compare with the black line in Fig.~\ref{fig:1}(b)] therefore indicating a strong suppression both of transverse and of longitudinal non-adiabatic effects. For carbon, the average ion energy per nucleon is $\varepsilon_p \approx 33.5$~MeV and the relative energy spread is $\Delta\varepsilon_p/\varepsilon_p \approx 0.06$. Notably, in addition to a substantial improvement of the quality of the carbon ion spectrum with respect to the single species foil case, here the obtained carbon energy per nucleon is larger than the prediction of the LS model ($\varepsilon_u \approx 21$~MeV). This occurs because, due to relativistic transparency, the laser pulse partially penetrates into the first layer, and consequently the mass of the foil that undergoes RPA acceleration is effectively smaller than that of the whole foil. 

Note that foils made of two distinct but intermingled ion species were proposed to suppress RT instability, with the heavy ion species lagging behind and absorbing most of the effect of RT instability~\cite{Tyu}. This is distinct from the case considered here, where suppression of transverse effects is essentially obtained by choosing short and intense laser pulses, and the two ions species are spatially separated, the first layer quenching longitudinal non-adiabatic effects and reducing the effective mass of the foil.

In conclusion, we have provided a method to suppress detrimental effects such as transverse instabilities and multiple ion population formation by employing short and intense laser pulses together with parameter matched double-layer foils. This allows to generate  ion beams that simultaneously exhibit high-energy per nucleon and a high-quality quasi-monochromatic spectrum. Finally, we mention that further increase of the ion energy per nucleon and simultaneous improvement of the monochromatic features of the ion spectrum are attainable with hybrid schemes where the first RPA stage is followed by a second stage of energy selection and eventually further acceleration with a compact linac~\cite{SSinigardi}.

\begin{acknowledgments}
The authors are grateful to Prof.~C.~H.~Keitel for his comments and his appreciation of this work. This article comprises parts of the PhD thesis work of Maitreyi Sangal to be submitted to the Heidelberg University, Germany.
\end{acknowledgments}

\bibliography{small}

\hyphenation{Post-Script Sprin-ger}\hyphenation{Post-Script Sprin-ger}
\begin{thebibliography}{51}%
\makeatletter
\providecommand \@ifxundefined [1]{%
 \@ifx{#1\undefined}
}%
\providecommand \@ifnum [1]{%
 \ifnum #1\expandafter \@firstoftwo
 \else \expandafter \@secondoftwo
 \fi
}%
\providecommand \@ifx [1]{%
 \ifx #1\expandafter \@firstoftwo
 \else \expandafter \@secondoftwo
 \fi
}%
\providecommand \natexlab [1]{#1}%
\providecommand \enquote  [1]{``#1''}%
\providecommand \bibnamefont  [1]{#1}%
\providecommand \bibfnamefont [1]{#1}%
\providecommand \citenamefont [1]{#1}%
\providecommand \href@noop [0]{\@secondoftwo}%
\providecommand \href [0]{\begingroup \@sanitize@url \@href}%
\providecommand \@href[1]{\@@startlink{#1}\@@href}%
\providecommand \@@href[1]{\endgroup#1\@@endlink}%
\providecommand \@sanitize@url [0]{\catcode `\\12\catcode `\$12\catcode
  `\&12\catcode `\#12\catcode `\^12\catcode `\_12\catcode `\%12\relax}%
\providecommand \@@startlink[1]{}%
\providecommand \@@endlink[0]{}%
\providecommand \url  [0]{\begingroup\@sanitize@url \@url }%
\providecommand \@url [1]{\endgroup\@href {#1}{\urlprefix }}%
\providecommand \urlprefix  [0]{URL }%
\providecommand \Eprint [0]{\href }%
\providecommand \doibase [0]{http://dx.doi.org/}%
\providecommand \selectlanguage [0]{\@gobble}%
\providecommand \bibinfo  [0]{\@secondoftwo}%
\providecommand \bibfield  [0]{\@secondoftwo}%
\providecommand \translation [1]{[#1]}%
\providecommand \BibitemOpen [0]{}%
\providecommand \bibitemStop [0]{}%
\providecommand \bibitemNoStop [0]{.\EOS\space}%
\providecommand \EOS [0]{\spacefactor3000\relax}%
\providecommand \BibitemShut  [1]{\csname bibitem#1\endcsname}%
\let\auto@bib@innerbib\@empty
\bibitem [{\citenamefont {Macchi}\ \emph {et~al.}(2013)\citenamefont {Macchi},
  \citenamefont {Borghesi},\ and\ \citenamefont {Passoni}}]{AMacchirev}%
  \BibitemOpen
  \bibfield  {author} {\bibinfo {author} {\bibfnamefont {A.}~\bibnamefont
  {Macchi}}, \bibinfo {author} {\bibfnamefont {M.}~\bibnamefont {Borghesi}}, \
  and\ \bibinfo {author} {\bibfnamefont {M.}~\bibnamefont {Passoni}},\ }\href
  {\doibase 10.1103/RevModPhys.85.751} {\bibfield  {journal} {\bibinfo
  {journal} {Rev. Mod. Phys.}\ }\textbf {\bibinfo {volume} {85}},\ \bibinfo
  {pages} {751} (\bibinfo {year} {2013})}\BibitemShut {NoStop}%
\bibitem [{\citenamefont {Daido}\ \emph {et~al.}(2012)\citenamefont {Daido},
  \citenamefont {Nishiuchi},\ and\ \citenamefont {Pirozhkov}}]{HDaido}%
  \BibitemOpen
  \bibfield  {author} {\bibinfo {author} {\bibfnamefont {H.}~\bibnamefont
  {Daido}}, \bibinfo {author} {\bibfnamefont {M.}~\bibnamefont {Nishiuchi}}, \
  and\ \bibinfo {author} {\bibfnamefont {A.~S.}\ \bibnamefont {Pirozhkov}},\
  }\href {\doibase 10.1088/0034-4885} {\bibfield  {journal} {\bibinfo
  {journal} {Rep. Prog. Phys.}\ }\textbf {\bibinfo {volume} {75}},\ \bibinfo
  {pages} {056401} (\bibinfo {year} {2012})}\BibitemShut {NoStop}%
\bibitem [{\citenamefont {Macchi}\ \emph {et~al.}(2009)\citenamefont {Macchi},
  \citenamefont {Veghini},\ and\ \citenamefont {Pegoraro}}]{AMacchiPRL09}%
  \BibitemOpen
  \bibfield  {author} {\bibinfo {author} {\bibfnamefont {A.}~\bibnamefont
  {Macchi}}, \bibinfo {author} {\bibfnamefont {S.}~\bibnamefont {Veghini}}, \
  and\ \bibinfo {author} {\bibfnamefont {F.}~\bibnamefont {Pegoraro}},\ }\href
  {\doibase 10.1103/PhysRevLett.103.085003} {\bibfield  {journal} {\bibinfo
  {journal} {Phys. Rev. Lett.}\ }\textbf {\bibinfo {volume} {103}},\ \bibinfo
  {pages} {085003} (\bibinfo {year} {2009})}\BibitemShut {NoStop}%
\bibitem [{\citenamefont {Macchi}\ \emph {et~al.}(2010)\citenamefont {Macchi},
  \citenamefont {Veghini}, \citenamefont {Liseykina},\ and\ \citenamefont
  {Pegoraro}}]{AMacchiRPA}%
  \BibitemOpen
  \bibfield  {author} {\bibinfo {author} {\bibfnamefont {A.}~\bibnamefont
  {Macchi}}, \bibinfo {author} {\bibfnamefont {S.}~\bibnamefont {Veghini}},
  \bibinfo {author} {\bibfnamefont {T.~V.}\ \bibnamefont {Liseykina}}, \ and\
  \bibinfo {author} {\bibfnamefont {F.}~\bibnamefont {Pegoraro}},\ }\href
  {http://stacks.iop.org/1367-2630/12/i=4/a=045013} {\bibfield  {journal}
  {\bibinfo  {journal} {New J. Phys.}\ }\textbf {\bibinfo {volume} {12}},\
  \bibinfo {pages} {045013} (\bibinfo {year} {2010})}\BibitemShut {NoStop}%
\bibitem [{\citenamefont {Tabak}\ \emph {et~al.}(1994)\citenamefont {Tabak},
  \citenamefont {Hammer}, \citenamefont {Glinsky}, \citenamefont {Kruer},
  \citenamefont {Wilks}, \citenamefont {Woodworth}, \citenamefont {Campbell},\
  and\ \citenamefont {Perry}}]{Tabak}%
  \BibitemOpen
  \bibfield  {author} {\bibinfo {author} {\bibfnamefont {M.}~\bibnamefont
  {Tabak}}, \bibinfo {author} {\bibfnamefont {J.}~\bibnamefont {Hammer}},
  \bibinfo {author} {\bibfnamefont {M.~E.}\ \bibnamefont {Glinsky}}, \bibinfo
  {author} {\bibfnamefont {W.~L.}\ \bibnamefont {Kruer}}, \bibinfo {author}
  {\bibfnamefont {S.~C.}\ \bibnamefont {Wilks}}, \bibinfo {author}
  {\bibfnamefont {J.}~\bibnamefont {Woodworth}}, \bibinfo {author}
  {\bibfnamefont {M.~E.}\ \bibnamefont {Campbell}}, \ and\ \bibinfo {author}
  {\bibfnamefont {M.~D.}\ \bibnamefont {Perry}},\ }\href {\doibase
  10.1063/1.870664} {\bibfield  {journal} {\bibinfo  {journal} {Phys. Plasmas}\
  }\textbf {\bibinfo {volume} {1}},\ \bibinfo {pages} {1626} (\bibinfo {year}
  {1994})}\BibitemShut {NoStop}%
\bibitem [{\citenamefont {Fern\'{a}ndez}\ \emph {et~al.}(2014)\citenamefont
  {Fern\'{a}ndez}, \citenamefont {Albright}, \citenamefont {Beg}, \citenamefont
  {Foord}, \citenamefont {Hegelich}, \citenamefont {Honrubia}, \citenamefont
  {Roth}, \citenamefont {Stephens},\ and\ \citenamefont {Yin}}]{JCFernandez}%
  \BibitemOpen
  \bibfield  {author} {\bibinfo {author} {\bibfnamefont {J.}~\bibnamefont
  {Fern\'{a}ndez}}, \bibinfo {author} {\bibfnamefont {B.}~\bibnamefont
  {Albright}}, \bibinfo {author} {\bibfnamefont {F.}~\bibnamefont {Beg}},
  \bibinfo {author} {\bibfnamefont {M.}~\bibnamefont {Foord}}, \bibinfo
  {author} {\bibfnamefont {B.}~\bibnamefont {Hegelich}}, \bibinfo {author}
  {\bibfnamefont {J.}~\bibnamefont {Honrubia}}, \bibinfo {author}
  {\bibfnamefont {M.}~\bibnamefont {Roth}}, \bibinfo {author} {\bibfnamefont
  {R.}~\bibnamefont {Stephens}}, \ and\ \bibinfo {author} {\bibfnamefont
  {L.}~\bibnamefont {Yin}},\ }\href
  {http://stacks.iop.org/0029-5515/54/i=5/a=054006} {\bibfield  {journal}
  {\bibinfo  {journal} {Nucl. Fusion}\ }\textbf {\bibinfo {volume} {54}},\
  \bibinfo {pages} {054006} (\bibinfo {year} {2014})}\BibitemShut {NoStop}%
\bibitem [{\citenamefont {Borghesi}\ \emph {et~al.}(2004)\citenamefont
  {Borghesi}, \citenamefont {Mackinnon}, \citenamefont {Campbell},
  \citenamefont {Hicks}, \citenamefont {Kar}, \citenamefont {Patel},
  \citenamefont {Price}, \citenamefont {Romagnani}, \citenamefont {Schiavi},\
  and\ \citenamefont {Willi}}]{BorghesiProbe}%
  \BibitemOpen
  \bibfield  {author} {\bibinfo {author} {\bibfnamefont {M.}~\bibnamefont
  {Borghesi}}, \bibinfo {author} {\bibfnamefont {A.~J.}\ \bibnamefont
  {Mackinnon}}, \bibinfo {author} {\bibfnamefont {D.~H.}\ \bibnamefont
  {Campbell}}, \bibinfo {author} {\bibfnamefont {D.~G.}\ \bibnamefont {Hicks}},
  \bibinfo {author} {\bibfnamefont {S.}~\bibnamefont {Kar}}, \bibinfo {author}
  {\bibfnamefont {P.~K.}\ \bibnamefont {Patel}}, \bibinfo {author}
  {\bibfnamefont {D.}~\bibnamefont {Price}}, \bibinfo {author} {\bibfnamefont
  {L.}~\bibnamefont {Romagnani}}, \bibinfo {author} {\bibfnamefont
  {A.}~\bibnamefont {Schiavi}}, \ and\ \bibinfo {author} {\bibfnamefont
  {O.}~\bibnamefont {Willi}},\ }\href {\doibase 10.1103/PhysRevLett.92.055003}
  {\bibfield  {journal} {\bibinfo  {journal} {Phys. Rev. Lett.}\ }\textbf
  {\bibinfo {volume} {92}},\ \bibinfo {pages} {055003} (\bibinfo {year}
  {2004})}\BibitemShut {NoStop}%
\bibitem [{\citenamefont {Borghesi}\ \emph {et~al.}(2008)\citenamefont
  {Borghesi}, \citenamefont {Bigongiari}, \citenamefont {Kar}, \citenamefont
  {Macchi}, \citenamefont {Romagnani}, \citenamefont {Audebert}, \citenamefont
  {Fuchs}, \citenamefont {Toncian}, \citenamefont {Willi}, \citenamefont
  {Bulanov}, \citenamefont {Mackinnon},\ and\ \citenamefont
  {Gauthier}}]{Borghesi_2008}%
  \BibitemOpen
  \bibfield  {author} {\bibinfo {author} {\bibfnamefont {M.}~\bibnamefont
  {Borghesi}}, \bibinfo {author} {\bibfnamefont {A.}~\bibnamefont
  {Bigongiari}}, \bibinfo {author} {\bibfnamefont {S.}~\bibnamefont {Kar}},
  \bibinfo {author} {\bibfnamefont {A.}~\bibnamefont {Macchi}}, \bibinfo
  {author} {\bibfnamefont {L.}~\bibnamefont {Romagnani}}, \bibinfo {author}
  {\bibfnamefont {P.}~\bibnamefont {Audebert}}, \bibinfo {author}
  {\bibfnamefont {J.}~\bibnamefont {Fuchs}}, \bibinfo {author} {\bibfnamefont
  {T.}~\bibnamefont {Toncian}}, \bibinfo {author} {\bibfnamefont
  {O.}~\bibnamefont {Willi}}, \bibinfo {author} {\bibfnamefont {S.~V.}\
  \bibnamefont {Bulanov}}, \bibinfo {author} {\bibfnamefont {A.~J.}\
  \bibnamefont {Mackinnon}}, \ and\ \bibinfo {author} {\bibfnamefont {J.~C.}\
  \bibnamefont {Gauthier}},\ }\href {\doibase 10.1088/0741-3335/50/12/124040}
  {\bibfield  {journal} {\bibinfo  {journal} {Plasma Phys. Control. Fusion}\
  }\textbf {\bibinfo {volume} {50}},\ \bibinfo {pages} {124040} (\bibinfo
  {year} {2008})}\BibitemShut {NoStop}%
\bibitem [{\citenamefont {Borghesi}\ \emph {et~al.}(2011)\citenamefont
  {Borghesi}, \citenamefont {Kar}, \citenamefont {Prasad}, \citenamefont
  {Kakolee}, \citenamefont {Quinn}, \citenamefont {Ahmed}, \citenamefont
  {Sarri}, \citenamefont {Ramakrishna}, \citenamefont {Qiao}, \citenamefont
  {Geissler}, \citenamefont {Ter-Avetisyan}, \citenamefont {Zepf},
  \citenamefont {Schettino}, \citenamefont {Stevens}, \citenamefont {Tolley},
  \citenamefont {Ward}, \citenamefont {Green}, \citenamefont {Foster},
  \citenamefont {Spindloe}, \citenamefont {Gallegos}, \citenamefont {Neely},
  \citenamefont {Carroll}, \citenamefont {Tresca}, \citenamefont {Yuan},
  \citenamefont {Quinn}, \citenamefont {McKenna}, \citenamefont {Dover},
  \citenamefont {Palmer}, \citenamefont {Schreiber}, \citenamefont {Najmudin},
  \citenamefont {Sari}, \citenamefont {Kraft}, \citenamefont {Merchant},
  \citenamefont {Jeynes}, \citenamefont {Kirkby}, \citenamefont {Fiorini},
  \citenamefont {Kirby},\ and\ \citenamefont {Green}}]{MBorghesiIBT}%
  \BibitemOpen
  \bibfield  {author} {\bibinfo {author} {\bibfnamefont {M.}~\bibnamefont
  {Borghesi}}, \bibinfo {author} {\bibfnamefont {S.}~\bibnamefont {Kar}},
  \bibinfo {author} {\bibfnamefont {R.}~\bibnamefont {Prasad}}, \bibinfo
  {author} {\bibfnamefont {F.~K.}\ \bibnamefont {Kakolee}}, \bibinfo {author}
  {\bibfnamefont {K.}~\bibnamefont {Quinn}}, \bibinfo {author} {\bibfnamefont
  {H.}~\bibnamefont {Ahmed}}, \bibinfo {author} {\bibfnamefont
  {G.}~\bibnamefont {Sarri}}, \bibinfo {author} {\bibfnamefont
  {B.}~\bibnamefont {Ramakrishna}}, \bibinfo {author} {\bibfnamefont
  {B.}~\bibnamefont {Qiao}}, \bibinfo {author} {\bibfnamefont {M.}~\bibnamefont
  {Geissler}}, \bibinfo {author} {\bibfnamefont {S.}~\bibnamefont
  {Ter-Avetisyan}}, \bibinfo {author} {\bibfnamefont {M.}~\bibnamefont {Zepf}},
  \bibinfo {author} {\bibfnamefont {G.}~\bibnamefont {Schettino}}, \bibinfo
  {author} {\bibfnamefont {B.}~\bibnamefont {Stevens}}, \bibinfo {author}
  {\bibfnamefont {M.}~\bibnamefont {Tolley}}, \bibinfo {author} {\bibfnamefont
  {A.}~\bibnamefont {Ward}}, \bibinfo {author} {\bibfnamefont {J.}~\bibnamefont
  {Green}}, \bibinfo {author} {\bibfnamefont {P.~S.}\ \bibnamefont {Foster}},
  \bibinfo {author} {\bibfnamefont {C.}~\bibnamefont {Spindloe}}, \bibinfo
  {author} {\bibfnamefont {A.}~\bibnamefont {Gallegos}, \bibfnamefont
  {P.~Robinson}}, \bibinfo {author} {\bibfnamefont {D.}~\bibnamefont {Neely}},
  \bibinfo {author} {\bibfnamefont {D.~C.}\ \bibnamefont {Carroll}}, \bibinfo
  {author} {\bibfnamefont {O.}~\bibnamefont {Tresca}}, \bibinfo {author}
  {\bibfnamefont {X.}~\bibnamefont {Yuan}}, \bibinfo {author} {\bibfnamefont
  {M.}~\bibnamefont {Quinn}}, \bibinfo {author} {\bibfnamefont
  {P.}~\bibnamefont {McKenna}}, \bibinfo {author} {\bibfnamefont
  {N.}~\bibnamefont {Dover}}, \bibinfo {author} {\bibfnamefont
  {C.}~\bibnamefont {Palmer}}, \bibinfo {author} {\bibfnamefont
  {J.}~\bibnamefont {Schreiber}}, \bibinfo {author} {\bibfnamefont
  {Z.}~\bibnamefont {Najmudin}}, \bibinfo {author} {\bibfnamefont
  {I.}~\bibnamefont {Sari}}, \bibinfo {author} {\bibfnamefont {M.}~\bibnamefont
  {Kraft}}, \bibinfo {author} {\bibfnamefont {M.}~\bibnamefont {Merchant}},
  \bibinfo {author} {\bibfnamefont {J.~C.}\ \bibnamefont {Jeynes}}, \bibinfo
  {author} {\bibfnamefont {K.}~\bibnamefont {Kirkby}}, \bibinfo {author}
  {\bibfnamefont {F.}~\bibnamefont {Fiorini}}, \bibinfo {author} {\bibfnamefont
  {D.}~\bibnamefont {Kirby}}, \ and\ \bibinfo {author} {\bibfnamefont
  {S.}~\bibnamefont {Green}}\ }(\bibinfo {year} {2011})\ pp.\ \bibinfo {pages}
  {171 -- 176}\BibitemShut {NoStop}%
\bibitem [{\citenamefont {Linz}\ and\ \citenamefont {Alonso}(2016)}]{ULinzRev}%
  \BibitemOpen
  \bibfield  {author} {\bibinfo {author} {\bibfnamefont {U.}~\bibnamefont
  {Linz}}\ and\ \bibinfo {author} {\bibfnamefont {J.}~\bibnamefont {Alonso}},\
  }\href {\doibase 10.1103/PhysRevAccelBeams.19.124802} {\bibfield  {journal}
  {\bibinfo  {journal} {Phys. Rev. Accel. Beams}\ }\textbf {\bibinfo {volume}
  {19}},\ \bibinfo {pages} {124802} (\bibinfo {year} {2016})}\BibitemShut
  {NoStop}%
\bibitem [{\citenamefont {Fritzler}\ \emph {et~al.}(2003)\citenamefont
  {Fritzler}, \citenamefont {Malka}, \citenamefont {Grillon}, \citenamefont
  {Rousseau}, \citenamefont {Burgy}, \citenamefont {Lefebvre}, \citenamefont
  {d’Humières}, \citenamefont {McKenna},\ and\ \citenamefont
  {Ledingham}}]{Fritzler}%
  \BibitemOpen
  \bibfield  {author} {\bibinfo {author} {\bibfnamefont {S.}~\bibnamefont
  {Fritzler}}, \bibinfo {author} {\bibfnamefont {V.}~\bibnamefont {Malka}},
  \bibinfo {author} {\bibfnamefont {G.}~\bibnamefont {Grillon}}, \bibinfo
  {author} {\bibfnamefont {J.}~\bibnamefont {Rousseau}}, \bibinfo {author}
  {\bibfnamefont {F.}~\bibnamefont {Burgy}}, \bibinfo {author} {\bibfnamefont
  {E.}~\bibnamefont {Lefebvre}}, \bibinfo {author} {\bibfnamefont
  {E.}~\bibnamefont {d’Humières}}, \bibinfo {author} {\bibfnamefont
  {P.}~\bibnamefont {McKenna}}, \ and\ \bibinfo {author} {\bibfnamefont
  {K.~W.~D.}\ \bibnamefont {Ledingham}},\ }\href {\doibase 10.1063/1.1616661}
  {\bibfield  {journal} {\bibinfo  {journal} {Appl. Phys. Lett.}\ }\textbf
  {\bibinfo {volume} {83}},\ \bibinfo {pages} {3039} (\bibinfo {year}
  {2003})}\BibitemShut {NoStop}%
\bibitem [{\citenamefont {McKenna}\ \emph {et~al.}(2004)\citenamefont
  {McKenna}, \citenamefont {Ledingham}, \citenamefont {Yang}, \citenamefont
  {Robson}, \citenamefont {McCanny}, \citenamefont {Shimizu}, \citenamefont
  {Clarke}, \citenamefont {Neely}, \citenamefont {Spohr}, \citenamefont
  {Chapman}, \citenamefont {Singhal}, \citenamefont {Krushelnick},
  \citenamefont {Wei},\ and\ \citenamefont {Norreys}}]{PMcKennaPRE}%
  \BibitemOpen
  \bibfield  {author} {\bibinfo {author} {\bibfnamefont {P.}~\bibnamefont
  {McKenna}}, \bibinfo {author} {\bibfnamefont {K.~W.~D.}\ \bibnamefont
  {Ledingham}}, \bibinfo {author} {\bibfnamefont {J.~M.}\ \bibnamefont {Yang}},
  \bibinfo {author} {\bibfnamefont {L.}~\bibnamefont {Robson}}, \bibinfo
  {author} {\bibfnamefont {T.}~\bibnamefont {McCanny}}, \bibinfo {author}
  {\bibfnamefont {S.}~\bibnamefont {Shimizu}}, \bibinfo {author} {\bibfnamefont
  {R.~J.}\ \bibnamefont {Clarke}}, \bibinfo {author} {\bibfnamefont
  {D.}~\bibnamefont {Neely}}, \bibinfo {author} {\bibfnamefont
  {K.}~\bibnamefont {Spohr}}, \bibinfo {author} {\bibfnamefont
  {R.}~\bibnamefont {Chapman}}, \bibinfo {author} {\bibfnamefont {R.~P.}\
  \bibnamefont {Singhal}}, \bibinfo {author} {\bibfnamefont {K.}~\bibnamefont
  {Krushelnick}}, \bibinfo {author} {\bibfnamefont {M.~S.}\ \bibnamefont
  {Wei}}, \ and\ \bibinfo {author} {\bibfnamefont {P.~A.}\ \bibnamefont
  {Norreys}},\ }\href {\doibase 10.1103/PhysRevE.70.036405} {\bibfield
  {journal} {\bibinfo  {journal} {Phys. Rev. E}\ }\textbf {\bibinfo {volume}
  {70}},\ \bibinfo {pages} {036405} (\bibinfo {year} {2004})}\BibitemShut
  {NoStop}%
\bibitem [{\citenamefont {Henig}\ \emph {et~al.}(2009)\citenamefont {Henig},
  \citenamefont {Steinke}, \citenamefont {Schn\"urer}, \citenamefont
  {Sokollik}, \citenamefont {H\"orlein}, \citenamefont {Kiefer}, \citenamefont
  {Jung}, \citenamefont {Schreiber}, \citenamefont {Hegelich}, \citenamefont
  {Yan}, \citenamefont {Meyer-ter Vehn}, \citenamefont {Tajima}, \citenamefont
  {Nickles}, \citenamefont {Sandner},\ and\ \citenamefont {Habs}}]{AHenig}%
  \BibitemOpen
  \bibfield  {author} {\bibinfo {author} {\bibfnamefont {A.}~\bibnamefont
  {Henig}}, \bibinfo {author} {\bibfnamefont {S.}~\bibnamefont {Steinke}},
  \bibinfo {author} {\bibfnamefont {M.}~\bibnamefont {Schn\"urer}}, \bibinfo
  {author} {\bibfnamefont {T.}~\bibnamefont {Sokollik}}, \bibinfo {author}
  {\bibfnamefont {R.}~\bibnamefont {H\"orlein}}, \bibinfo {author}
  {\bibfnamefont {D.}~\bibnamefont {Kiefer}}, \bibinfo {author} {\bibfnamefont
  {D.}~\bibnamefont {Jung}}, \bibinfo {author} {\bibfnamefont {J.}~\bibnamefont
  {Schreiber}}, \bibinfo {author} {\bibfnamefont {B.~M.}\ \bibnamefont
  {Hegelich}}, \bibinfo {author} {\bibfnamefont {X.~Q.}\ \bibnamefont {Yan}},
  \bibinfo {author} {\bibfnamefont {J.}~\bibnamefont {Meyer-ter Vehn}},
  \bibinfo {author} {\bibfnamefont {T.}~\bibnamefont {Tajima}}, \bibinfo
  {author} {\bibfnamefont {P.~V.}\ \bibnamefont {Nickles}}, \bibinfo {author}
  {\bibfnamefont {W.}~\bibnamefont {Sandner}}, \ and\ \bibinfo {author}
  {\bibfnamefont {D.}~\bibnamefont {Habs}},\ }\href {\doibase
  10.1103/PhysRevLett.103.245003} {\bibfield  {journal} {\bibinfo  {journal}
  {Phys. Rev. Lett.}\ }\textbf {\bibinfo {volume} {103}},\ \bibinfo {pages}
  {245003} (\bibinfo {year} {2009})}\BibitemShut {NoStop}%
\bibitem [{\citenamefont {Kar}\ \emph {et~al.}(2012)\citenamefont {Kar},
  \citenamefont {Kakolee}, \citenamefont {Qiao}, \citenamefont {Macchi},
  \citenamefont {Cerchez}, \citenamefont {Doria}, \citenamefont {Geissler},
  \citenamefont {McKenna}, \citenamefont {Neely}, \citenamefont {Osterholz},
  \citenamefont {Prasad}, \citenamefont {Quinn}, \citenamefont {Ramakrishna},
  \citenamefont {Sarri}, \citenamefont {Willi}, \citenamefont {Yuan},
  \citenamefont {Zepf},\ and\ \citenamefont {Borghesi}}]{SKarMultsp}%
  \BibitemOpen
  \bibfield  {author} {\bibinfo {author} {\bibfnamefont {S.}~\bibnamefont
  {Kar}}, \bibinfo {author} {\bibfnamefont {K.~F.}\ \bibnamefont {Kakolee}},
  \bibinfo {author} {\bibfnamefont {B.}~\bibnamefont {Qiao}}, \bibinfo {author}
  {\bibfnamefont {A.}~\bibnamefont {Macchi}}, \bibinfo {author} {\bibfnamefont
  {M.}~\bibnamefont {Cerchez}}, \bibinfo {author} {\bibfnamefont
  {D.}~\bibnamefont {Doria}}, \bibinfo {author} {\bibfnamefont
  {M.}~\bibnamefont {Geissler}}, \bibinfo {author} {\bibfnamefont
  {P.}~\bibnamefont {McKenna}}, \bibinfo {author} {\bibfnamefont
  {D.}~\bibnamefont {Neely}}, \bibinfo {author} {\bibfnamefont
  {J.}~\bibnamefont {Osterholz}}, \bibinfo {author} {\bibfnamefont
  {R.}~\bibnamefont {Prasad}}, \bibinfo {author} {\bibfnamefont
  {K.}~\bibnamefont {Quinn}}, \bibinfo {author} {\bibfnamefont
  {B.}~\bibnamefont {Ramakrishna}}, \bibinfo {author} {\bibfnamefont
  {G.}~\bibnamefont {Sarri}}, \bibinfo {author} {\bibfnamefont
  {O.}~\bibnamefont {Willi}}, \bibinfo {author} {\bibfnamefont {X.~Y.}\
  \bibnamefont {Yuan}}, \bibinfo {author} {\bibfnamefont {M.}~\bibnamefont
  {Zepf}}, \ and\ \bibinfo {author} {\bibfnamefont {M.}~\bibnamefont
  {Borghesi}},\ }\href {\doibase 10.1103/PhysRevLett.109.185006} {\bibfield
  {journal} {\bibinfo  {journal} {Phys. Rev. Lett.}\ }\textbf {\bibinfo
  {volume} {109}},\ \bibinfo {pages} {185006} (\bibinfo {year}
  {2012})}\BibitemShut {NoStop}%
\bibitem [{\citenamefont {Aurand}\ \emph {et~al.}(2013)\citenamefont {Aurand},
  \citenamefont {Kuschel}, \citenamefont {J\"{a}ckel}, \citenamefont
  {R\"{o}del}, \citenamefont {Zhao}, \citenamefont {Herzer}, \citenamefont
  {Paz}, \citenamefont {Bierbach}, \citenamefont {Polz}, \citenamefont {Elkin},
  \citenamefont {Paulus}, \citenamefont {Karmakar}, \citenamefont {Gibbon},
  \citenamefont {Kuehl},\ and\ \citenamefont {Kaluza}}]{Aurand_2013}%
  \BibitemOpen
  \bibfield  {author} {\bibinfo {author} {\bibfnamefont {B.}~\bibnamefont
  {Aurand}}, \bibinfo {author} {\bibfnamefont {S.}~\bibnamefont {Kuschel}},
  \bibinfo {author} {\bibfnamefont {O.}~\bibnamefont {J\"{a}ckel}}, \bibinfo
  {author} {\bibfnamefont {C.}~\bibnamefont {R\"{o}del}}, \bibinfo {author}
  {\bibfnamefont {H.~Y.}\ \bibnamefont {Zhao}}, \bibinfo {author}
  {\bibfnamefont {S.}~\bibnamefont {Herzer}}, \bibinfo {author} {\bibfnamefont
  {A.~E.}\ \bibnamefont {Paz}}, \bibinfo {author} {\bibfnamefont
  {J.}~\bibnamefont {Bierbach}}, \bibinfo {author} {\bibfnamefont
  {J.}~\bibnamefont {Polz}}, \bibinfo {author} {\bibfnamefont {B.}~\bibnamefont
  {Elkin}}, \bibinfo {author} {\bibfnamefont {G.~G.}\ \bibnamefont {Paulus}},
  \bibinfo {author} {\bibfnamefont {A.}~\bibnamefont {Karmakar}}, \bibinfo
  {author} {\bibfnamefont {P.}~\bibnamefont {Gibbon}}, \bibinfo {author}
  {\bibfnamefont {T.}~\bibnamefont {Kuehl}}, \ and\ \bibinfo {author}
  {\bibfnamefont {M.~C.}\ \bibnamefont {Kaluza}},\ }\href {\doibase
  10.1088/1367-2630/15/3/033031} {\bibfield  {journal} {\bibinfo  {journal}
  {New J. Phys.}\ }\textbf {\bibinfo {volume} {15}},\ \bibinfo {pages} {033031}
  (\bibinfo {year} {2013})}\BibitemShut {NoStop}%
\bibitem [{\citenamefont {Palmer}\ \emph {et~al.}(2012)\citenamefont {Palmer},
  \citenamefont {Schreiber}, \citenamefont {Nagel}, \citenamefont {Dover},
  \citenamefont {Bellei}, \citenamefont {Beg}, \citenamefont {Bott},
  \citenamefont {Clarke}, \citenamefont {Dangor}, \citenamefont {Hassan},
  \citenamefont {Hilz}, \citenamefont {Jung}, \citenamefont {Kneip},
  \citenamefont {Mangles}, \citenamefont {Lancaster}, \citenamefont {Rehman},
  \citenamefont {Robinson}, \citenamefont {Spindloe}, \citenamefont {Szerypo},
  \citenamefont {Tatarakis}, \citenamefont {Yeung}, \citenamefont {Zepf},\ and\
  \citenamefont {Najmudin}}]{CAJPalmerRTIExp}%
  \BibitemOpen
  \bibfield  {author} {\bibinfo {author} {\bibfnamefont {C.~A.~J.}\
  \bibnamefont {Palmer}}, \bibinfo {author} {\bibfnamefont {J.}~\bibnamefont
  {Schreiber}}, \bibinfo {author} {\bibfnamefont {S.~R.}\ \bibnamefont
  {Nagel}}, \bibinfo {author} {\bibfnamefont {N.~P.}\ \bibnamefont {Dover}},
  \bibinfo {author} {\bibfnamefont {C.}~\bibnamefont {Bellei}}, \bibinfo
  {author} {\bibfnamefont {F.~N.}\ \bibnamefont {Beg}}, \bibinfo {author}
  {\bibfnamefont {S.}~\bibnamefont {Bott}}, \bibinfo {author} {\bibfnamefont
  {R.~J.}\ \bibnamefont {Clarke}}, \bibinfo {author} {\bibfnamefont {A.~E.}\
  \bibnamefont {Dangor}}, \bibinfo {author} {\bibfnamefont {S.~M.}\
  \bibnamefont {Hassan}}, \bibinfo {author} {\bibfnamefont {P.}~\bibnamefont
  {Hilz}}, \bibinfo {author} {\bibfnamefont {D.}~\bibnamefont {Jung}}, \bibinfo
  {author} {\bibfnamefont {S.}~\bibnamefont {Kneip}}, \bibinfo {author}
  {\bibfnamefont {S.~P.~D.}\ \bibnamefont {Mangles}}, \bibinfo {author}
  {\bibfnamefont {K.~L.}\ \bibnamefont {Lancaster}}, \bibinfo {author}
  {\bibfnamefont {A.}~\bibnamefont {Rehman}}, \bibinfo {author} {\bibfnamefont
  {A.~P.~L.}\ \bibnamefont {Robinson}}, \bibinfo {author} {\bibfnamefont
  {C.}~\bibnamefont {Spindloe}}, \bibinfo {author} {\bibfnamefont
  {J.}~\bibnamefont {Szerypo}}, \bibinfo {author} {\bibfnamefont
  {M.}~\bibnamefont {Tatarakis}}, \bibinfo {author} {\bibfnamefont
  {M.}~\bibnamefont {Yeung}}, \bibinfo {author} {\bibfnamefont
  {M.}~\bibnamefont {Zepf}}, \ and\ \bibinfo {author} {\bibfnamefont
  {Z.}~\bibnamefont {Najmudin}},\ }\href {\doibase
  10.1103/PhysRevLett.108.225002} {\bibfield  {journal} {\bibinfo  {journal}
  {Phys. Rev. Lett.}\ }\textbf {\bibinfo {volume} {108}},\ \bibinfo {pages}
  {225002} (\bibinfo {year} {2012})}\BibitemShut {NoStop}%
\bibitem [{\citenamefont {Steinke}\ \emph {et~al.}(2013)\citenamefont
  {Steinke}, \citenamefont {Hilz}, \citenamefont {Schn\"urer}, \citenamefont
  {Priebe}, \citenamefont {Br\"anzel}, \citenamefont {Abicht}, \citenamefont
  {Kiefer}, \citenamefont {Kreuzer}, \citenamefont {Ostermayr}, \citenamefont
  {Schreiber}, \citenamefont {Andreev}, \citenamefont {Yu}, \citenamefont
  {Pukhov},\ and\ \citenamefont {Sandner}}]{SsteinkeStable}%
  \BibitemOpen
  \bibfield  {author} {\bibinfo {author} {\bibfnamefont {S.}~\bibnamefont
  {Steinke}}, \bibinfo {author} {\bibfnamefont {P.}~\bibnamefont {Hilz}},
  \bibinfo {author} {\bibfnamefont {M.}~\bibnamefont {Schn\"urer}}, \bibinfo
  {author} {\bibfnamefont {G.}~\bibnamefont {Priebe}}, \bibinfo {author}
  {\bibfnamefont {J.}~\bibnamefont {Br\"anzel}}, \bibinfo {author}
  {\bibfnamefont {F.}~\bibnamefont {Abicht}}, \bibinfo {author} {\bibfnamefont
  {D.}~\bibnamefont {Kiefer}}, \bibinfo {author} {\bibfnamefont
  {C.}~\bibnamefont {Kreuzer}}, \bibinfo {author} {\bibfnamefont
  {T.}~\bibnamefont {Ostermayr}}, \bibinfo {author} {\bibfnamefont
  {J.}~\bibnamefont {Schreiber}}, \bibinfo {author} {\bibfnamefont {A.~A.}\
  \bibnamefont {Andreev}}, \bibinfo {author} {\bibfnamefont {T.~P.}\
  \bibnamefont {Yu}}, \bibinfo {author} {\bibfnamefont {A.}~\bibnamefont
  {Pukhov}}, \ and\ \bibinfo {author} {\bibfnamefont {W.}~\bibnamefont
  {Sandner}},\ }\href {\doibase 10.1103/PhysRevSTAB.16.011303} {\bibfield
  {journal} {\bibinfo  {journal} {Phys. Rev. ST Accel. Beams}\ }\textbf
  {\bibinfo {volume} {16}},\ \bibinfo {pages} {011303} (\bibinfo {year}
  {2013})}\BibitemShut {NoStop}%
\bibitem [{\citenamefont {Bin}\ \emph {et~al.}(2015)\citenamefont {Bin},
  \citenamefont {Ma}, \citenamefont {Wang}, \citenamefont {Streeter},
  \citenamefont {Kreuzer}, \citenamefont {Kiefer}, \citenamefont {Yeung},
  \citenamefont {Cousens}, \citenamefont {Foster}, \citenamefont {Dromey},
  \citenamefont {Yan}, \citenamefont {Ramis}, \citenamefont {Meyer-ter Vehn},
  \citenamefont {Zepf},\ and\ \citenamefont {Schreiber}}]{JHBin}%
  \BibitemOpen
  \bibfield  {author} {\bibinfo {author} {\bibfnamefont {J.~H.}\ \bibnamefont
  {Bin}}, \bibinfo {author} {\bibfnamefont {W.~J.}\ \bibnamefont {Ma}},
  \bibinfo {author} {\bibfnamefont {H.~Y.}\ \bibnamefont {Wang}}, \bibinfo
  {author} {\bibfnamefont {M.~J.~V.}\ \bibnamefont {Streeter}}, \bibinfo
  {author} {\bibfnamefont {C.}~\bibnamefont {Kreuzer}}, \bibinfo {author}
  {\bibfnamefont {D.}~\bibnamefont {Kiefer}}, \bibinfo {author} {\bibfnamefont
  {M.}~\bibnamefont {Yeung}}, \bibinfo {author} {\bibfnamefont
  {S.}~\bibnamefont {Cousens}}, \bibinfo {author} {\bibfnamefont {P.~S.}\
  \bibnamefont {Foster}}, \bibinfo {author} {\bibfnamefont {B.}~\bibnamefont
  {Dromey}}, \bibinfo {author} {\bibfnamefont {X.~Q.}\ \bibnamefont {Yan}},
  \bibinfo {author} {\bibfnamefont {R.}~\bibnamefont {Ramis}}, \bibinfo
  {author} {\bibfnamefont {J.}~\bibnamefont {Meyer-ter Vehn}}, \bibinfo
  {author} {\bibfnamefont {M.}~\bibnamefont {Zepf}}, \ and\ \bibinfo {author}
  {\bibfnamefont {J.}~\bibnamefont {Schreiber}},\ }\href {\doibase
  10.1103/PhysRevLett.115.064801} {\bibfield  {journal} {\bibinfo  {journal}
  {Phys. Rev. Lett.}\ }\textbf {\bibinfo {volume} {115}},\ \bibinfo {pages}
  {064801} (\bibinfo {year} {2015})}\BibitemShut {NoStop}%
\bibitem [{Eli(2014)}]{EliURL}%
  \BibitemOpen
  \href {http://www.eli-laser.eu/} {\enquote {\bibinfo {title} {Extreme light
  infrastructure},}\ }\bibinfo {howpublished} {\url{http://www.eli-laser.eu/}}
  (\bibinfo {year} {2014})\BibitemShut {NoStop}%
\bibitem [{Apo(2014)}]{ApollonURL}%
  \BibitemOpen
  \href {https://portail.polytechnique.edu/luli/en/cilex-apollon/apollon}
  {\enquote {\bibinfo {title} {Apollon},}\ }\bibinfo {howpublished}
  {\url{https://portail.polytechnique.edu/luli/en/cilex-apollon/apollon}}
  (\bibinfo {year} {2014})\BibitemShut {NoStop}%
\bibitem [{\citenamefont {{Ch{\'e}riaux}}\ \emph {et~al.}(2012)\citenamefont
  {{Ch{\'e}riaux}}, \citenamefont {{Giambruno}}, \citenamefont
  {{Fr{\'e}neaux}}, \citenamefont {{Leconte}}, \citenamefont {{Ramirez}},
  \citenamefont {{Georges}}, \citenamefont {{Druon}}, \citenamefont
  {{Papadopoulos}}, \citenamefont {{Pellegrina}}, \citenamefont {{Le Blanc}},
  \citenamefont {{Doyen}}, \citenamefont {{Legat}}, \citenamefont {{Boudenne}},
  \citenamefont {{Mennerat}}, \citenamefont {{Audebert}}, \citenamefont
  {{Mourou}}, \citenamefont {{Mathieu}},\ and\ \citenamefont
  {{Chambaret}}}]{apollon}%
  \BibitemOpen
  \bibfield  {author} {\bibinfo {author} {\bibfnamefont {G.}~\bibnamefont
  {{Ch{\'e}riaux}}}, \bibinfo {author} {\bibfnamefont {F.}~\bibnamefont
  {{Giambruno}}}, \bibinfo {author} {\bibfnamefont {A.}~\bibnamefont
  {{Fr{\'e}neaux}}}, \bibinfo {author} {\bibfnamefont {F.}~\bibnamefont
  {{Leconte}}}, \bibinfo {author} {\bibfnamefont {L.~P.}\ \bibnamefont
  {{Ramirez}}}, \bibinfo {author} {\bibfnamefont {P.}~\bibnamefont
  {{Georges}}}, \bibinfo {author} {\bibfnamefont {F.}~\bibnamefont {{Druon}}},
  \bibinfo {author} {\bibfnamefont {D.~N.}\ \bibnamefont {{Papadopoulos}}},
  \bibinfo {author} {\bibfnamefont {A.}~\bibnamefont {{Pellegrina}}}, \bibinfo
  {author} {\bibfnamefont {C.}~\bibnamefont {{Le Blanc}}}, \bibinfo {author}
  {\bibfnamefont {I.}~\bibnamefont {{Doyen}}}, \bibinfo {author} {\bibfnamefont
  {L.}~\bibnamefont {{Legat}}}, \bibinfo {author} {\bibfnamefont {J.~M.}\
  \bibnamefont {{Boudenne}}}, \bibinfo {author} {\bibfnamefont
  {G.}~\bibnamefont {{Mennerat}}}, \bibinfo {author} {\bibfnamefont
  {P.}~\bibnamefont {{Audebert}}}, \bibinfo {author} {\bibfnamefont
  {G.}~\bibnamefont {{Mourou}}}, \bibinfo {author} {\bibfnamefont
  {F.}~\bibnamefont {{Mathieu}}}, \ and\ \bibinfo {author} {\bibfnamefont
  {J.~P.}\ \bibnamefont {{Chambaret}}},\ }\href {\doibase 10.1063/1.4736764}
  {\bibfield  {journal} {\bibinfo  {journal} {AIP Conf. Proc.}\ }\textbf
  {\bibinfo {volume} {1462}},\ \bibinfo {pages} {78} (\bibinfo {year}
  {2012})}\BibitemShut {NoStop}%
\bibitem [{xce(2014)}]{xcelsURL}%
  \BibitemOpen
  \href {http://www.xcels.iapras.ru/} {\enquote {\bibinfo {title} {Exawatt
  center for extreme light studies},}\ }\bibinfo {howpublished}
  {\url{https://xcels.iapras.ru/}} (\bibinfo {year} {2014})\BibitemShut
  {NoStop}%
\bibitem [{\citenamefont {Pegoraro}\ and\ \citenamefont
  {Bulanov}(2007)}]{FPegoraroBubbles}%
  \BibitemOpen
  \bibfield  {author} {\bibinfo {author} {\bibfnamefont {F.}~\bibnamefont
  {Pegoraro}}\ and\ \bibinfo {author} {\bibfnamefont {S.~V.}\ \bibnamefont
  {Bulanov}},\ }\href {\doibase 10.1103/PhysRevLett.99.065002} {\bibfield
  {journal} {\bibinfo  {journal} {Phys. Rev. Lett.}\ }\textbf {\bibinfo
  {volume} {99}},\ \bibinfo {pages} {065002} (\bibinfo {year}
  {2007})}\BibitemShut {NoStop}%
\bibitem [{\citenamefont {Zhang}\ \emph {et~al.}(2011)\citenamefont {Zhang},
  \citenamefont {Shen}, \citenamefont {Ji}, \citenamefont {Wang}, \citenamefont
  {Xu}, \citenamefont {Yu},\ and\ \citenamefont {Wang}}]{ZXiaomei}%
  \BibitemOpen
  \bibfield  {author} {\bibinfo {author} {\bibfnamefont {X.}~\bibnamefont
  {Zhang}}, \bibinfo {author} {\bibfnamefont {B.}~\bibnamefont {Shen}},
  \bibinfo {author} {\bibfnamefont {L.}~\bibnamefont {Ji}}, \bibinfo {author}
  {\bibfnamefont {W.}~\bibnamefont {Wang}}, \bibinfo {author} {\bibfnamefont
  {J.}~\bibnamefont {Xu}}, \bibinfo {author} {\bibfnamefont {Y.}~\bibnamefont
  {Yu}}, \ and\ \bibinfo {author} {\bibfnamefont {X.}~\bibnamefont {Wang}},\
  }\href {\doibase 10.1063/1.3603821} {\bibfield  {journal} {\bibinfo
  {journal} {Phys. Plasmas}\ }\textbf {\bibinfo {volume} {18}},\ \bibinfo
  {pages} {073101} (\bibinfo {year} {2011})}\BibitemShut {NoStop}%
\bibitem [{\citenamefont {Eliasson}(2015)}]{Eliasson2015}%
  \BibitemOpen
  \bibfield  {author} {\bibinfo {author} {\bibfnamefont {B.}~\bibnamefont
  {Eliasson}},\ }\href {\doibase 10.1088/1367-2630/17/3/033026} {\bibfield
  {journal} {\bibinfo  {journal} {New Journal of Physics}\ }\textbf {\bibinfo
  {volume} {17}},\ \bibinfo {pages} {033026} (\bibinfo {year}
  {2015})}\BibitemShut {NoStop}%
\bibitem [{\citenamefont {Wan}\ \emph {et~al.}(2016)\citenamefont {Wan},
  \citenamefont {Pai}, \citenamefont {Zhang}, \citenamefont {Li}, \citenamefont
  {Wu}, \citenamefont {Hua}, \citenamefont {Lu}, \citenamefont {Gu},
  \citenamefont {Silva}, \citenamefont {Joshi},\ and\ \citenamefont
  {Mori}}]{YWanPhysMech}%
  \BibitemOpen
  \bibfield  {author} {\bibinfo {author} {\bibfnamefont {Y.}~\bibnamefont
  {Wan}}, \bibinfo {author} {\bibfnamefont {C.-H.}\ \bibnamefont {Pai}},
  \bibinfo {author} {\bibfnamefont {C.~J.}\ \bibnamefont {Zhang}}, \bibinfo
  {author} {\bibfnamefont {F.}~\bibnamefont {Li}}, \bibinfo {author}
  {\bibfnamefont {Y.~P.}\ \bibnamefont {Wu}}, \bibinfo {author} {\bibfnamefont
  {J.~F.}\ \bibnamefont {Hua}}, \bibinfo {author} {\bibfnamefont
  {W.}~\bibnamefont {Lu}}, \bibinfo {author} {\bibfnamefont {Y.~Q.}\
  \bibnamefont {Gu}}, \bibinfo {author} {\bibfnamefont {L.~O.}\ \bibnamefont
  {Silva}}, \bibinfo {author} {\bibfnamefont {C.}~\bibnamefont {Joshi}}, \ and\
  \bibinfo {author} {\bibfnamefont {W.~B.}\ \bibnamefont {Mori}},\ }\href
  {\doibase 10.1103/PhysRevLett.117.234801} {\bibfield  {journal} {\bibinfo
  {journal} {Phys. Rev. Lett.}\ }\textbf {\bibinfo {volume} {117}},\ \bibinfo
  {pages} {234801} (\bibinfo {year} {2016})}\BibitemShut {NoStop}%
\bibitem [{\citenamefont {Dollar}\ \emph {et~al.}(2012)\citenamefont {Dollar},
  \citenamefont {Zulick}, \citenamefont {Thomas}, \citenamefont {Chvykov},
  \citenamefont {Davis}, \citenamefont {Kalinchenko}, \citenamefont {Matsuoka},
  \citenamefont {McGuffey}, \citenamefont {Petrov}, \citenamefont {Willingale},
  \citenamefont {Yanovsky}, \citenamefont {Maksimchuk},\ and\ \citenamefont
  {Krushelnick}}]{FDollar}%
  \BibitemOpen
  \bibfield  {author} {\bibinfo {author} {\bibfnamefont {F.}~\bibnamefont
  {Dollar}}, \bibinfo {author} {\bibfnamefont {C.}~\bibnamefont {Zulick}},
  \bibinfo {author} {\bibfnamefont {A.~G.~R.}\ \bibnamefont {Thomas}}, \bibinfo
  {author} {\bibfnamefont {V.}~\bibnamefont {Chvykov}}, \bibinfo {author}
  {\bibfnamefont {J.}~\bibnamefont {Davis}}, \bibinfo {author} {\bibfnamefont
  {G.}~\bibnamefont {Kalinchenko}}, \bibinfo {author} {\bibfnamefont
  {T.}~\bibnamefont {Matsuoka}}, \bibinfo {author} {\bibfnamefont
  {C.}~\bibnamefont {McGuffey}}, \bibinfo {author} {\bibfnamefont {G.~M.}\
  \bibnamefont {Petrov}}, \bibinfo {author} {\bibfnamefont {L.}~\bibnamefont
  {Willingale}}, \bibinfo {author} {\bibfnamefont {V.}~\bibnamefont
  {Yanovsky}}, \bibinfo {author} {\bibfnamefont {A.}~\bibnamefont
  {Maksimchuk}}, \ and\ \bibinfo {author} {\bibfnamefont {K.}~\bibnamefont
  {Krushelnick}},\ }\href {\doibase 10.1103/PhysRevLett.108.175005} {\bibfield
  {journal} {\bibinfo  {journal} {Phys. Rev. Lett.}\ }\textbf {\bibinfo
  {volume} {108}},\ \bibinfo {pages} {175005} (\bibinfo {year}
  {2012})}\BibitemShut {NoStop}%
\bibitem [{\citenamefont {Bulanov}\ \emph {et~al.}(2010)\citenamefont
  {Bulanov}, \citenamefont {Echkina}, \citenamefont {Esirkepov}, \citenamefont
  {Inovenkov}, \citenamefont {Kando}, \citenamefont {Pegoraro},\ and\
  \citenamefont {Korn}}]{SVBulanov}%
  \BibitemOpen
  \bibfield  {author} {\bibinfo {author} {\bibfnamefont {S.~V.}\ \bibnamefont
  {Bulanov}}, \bibinfo {author} {\bibfnamefont {E.~Y.}\ \bibnamefont
  {Echkina}}, \bibinfo {author} {\bibfnamefont {T.~Z.}\ \bibnamefont
  {Esirkepov}}, \bibinfo {author} {\bibfnamefont {I.~N.}\ \bibnamefont
  {Inovenkov}}, \bibinfo {author} {\bibfnamefont {M.}~\bibnamefont {Kando}},
  \bibinfo {author} {\bibfnamefont {F.}~\bibnamefont {Pegoraro}}, \ and\
  \bibinfo {author} {\bibfnamefont {G.}~\bibnamefont {Korn}},\ }\href {\doibase
  10.1103/PhysRevLett.104.135003} {\bibfield  {journal} {\bibinfo  {journal}
  {Phys. Rev. Lett.}\ }\textbf {\bibinfo {volume} {104}},\ \bibinfo {pages}
  {135003} (\bibinfo {year} {2010})}\BibitemShut {NoStop}%
\bibitem [{\citenamefont {Tamburini}\ \emph {et~al.}(2012)\citenamefont
  {Tamburini}, \citenamefont {Liseykina}, \citenamefont {Pegoraro},\ and\
  \citenamefont {Macchi}}]{MTamburiniRR}%
  \BibitemOpen
  \bibfield  {author} {\bibinfo {author} {\bibfnamefont {M.}~\bibnamefont
  {Tamburini}}, \bibinfo {author} {\bibfnamefont {T.~V.}\ \bibnamefont
  {Liseykina}}, \bibinfo {author} {\bibfnamefont {F.}~\bibnamefont {Pegoraro}},
  \ and\ \bibinfo {author} {\bibfnamefont {A.}~\bibnamefont {Macchi}},\ }\href
  {\doibase 10.1103/PhysRevE.85.016407} {\bibfield  {journal} {\bibinfo
  {journal} {Phys. Rev. E}\ }\textbf {\bibinfo {volume} {85}},\ \bibinfo
  {pages} {016407} (\bibinfo {year} {2012})}\BibitemShut {NoStop}%
\bibitem [{\citenamefont {Bulanov}\ \emph {et~al.}(2012)\citenamefont
  {Bulanov}, \citenamefont {Schroeder}, \citenamefont {Esarey},\ and\
  \citenamefont {Leemans}}]{SSBulanov}%
  \BibitemOpen
  \bibfield  {author} {\bibinfo {author} {\bibfnamefont {S.~S.}\ \bibnamefont
  {Bulanov}}, \bibinfo {author} {\bibfnamefont {C.~B.}\ \bibnamefont
  {Schroeder}}, \bibinfo {author} {\bibfnamefont {E.}~\bibnamefont {Esarey}}, \
  and\ \bibinfo {author} {\bibfnamefont {W.~P.}\ \bibnamefont {Leemans}},\
  }\href {\doibase 10.1063/1.4752214} {\bibfield  {journal} {\bibinfo
  {journal} {Phys. Plasmas}\ }\textbf {\bibinfo {volume} {19}},\ \bibinfo
  {pages} {093112} (\bibinfo {year} {2012})}\BibitemShut {NoStop}%
\bibitem [{\citenamefont {Bulanov}\ \emph {et~al.}(2015)\citenamefont
  {Bulanov}, \citenamefont {Esarey}, \citenamefont {Schroeder}, \citenamefont
  {Bulanov}, \citenamefont {Esirkepov}, \citenamefont {Kando}, \citenamefont
  {Pegoraro},\ and\ \citenamefont {Leemans}}]{BulanovGuid}%
  \BibitemOpen
  \bibfield  {author} {\bibinfo {author} {\bibfnamefont {S.~S.}\ \bibnamefont
  {Bulanov}}, \bibinfo {author} {\bibfnamefont {E.}~\bibnamefont {Esarey}},
  \bibinfo {author} {\bibfnamefont {C.~B.}\ \bibnamefont {Schroeder}}, \bibinfo
  {author} {\bibfnamefont {S.~V.}\ \bibnamefont {Bulanov}}, \bibinfo {author}
  {\bibfnamefont {T.~Z.}\ \bibnamefont {Esirkepov}}, \bibinfo {author}
  {\bibfnamefont {M.}~\bibnamefont {Kando}}, \bibinfo {author} {\bibfnamefont
  {F.}~\bibnamefont {Pegoraro}}, \ and\ \bibinfo {author} {\bibfnamefont
  {W.~P.}\ \bibnamefont {Leemans}},\ }\href {\doibase
  10.1103/PhysRevLett.114.105003} {\bibfield  {journal} {\bibinfo  {journal}
  {Phys. Rev. Lett.}\ }\textbf {\bibinfo {volume} {114}},\ \bibinfo {pages}
  {105003} (\bibinfo {year} {2015})}\BibitemShut {NoStop}%
\bibitem [{\citenamefont {Bulanov}\ \emph {et~al.}(2008)\citenamefont
  {Bulanov}, \citenamefont {Brantov}, \citenamefont {Bychenkov}, \citenamefont
  {Chvykov}, \citenamefont {Kalinchenko}, \citenamefont {Matsuoka},
  \citenamefont {Rousseau}, \citenamefont {Reed}, \citenamefont {Yanovsky},
  \citenamefont {Litzenberg}, \citenamefont {Krushelnick},\ and\ \citenamefont
  {Maksimchuk}}]{BulanovFlattop}%
  \BibitemOpen
  \bibfield  {author} {\bibinfo {author} {\bibfnamefont {S.~S.}\ \bibnamefont
  {Bulanov}}, \bibinfo {author} {\bibfnamefont {A.}~\bibnamefont {Brantov}},
  \bibinfo {author} {\bibfnamefont {V.~Y.}\ \bibnamefont {Bychenkov}}, \bibinfo
  {author} {\bibfnamefont {V.}~\bibnamefont {Chvykov}}, \bibinfo {author}
  {\bibfnamefont {G.}~\bibnamefont {Kalinchenko}}, \bibinfo {author}
  {\bibfnamefont {T.}~\bibnamefont {Matsuoka}}, \bibinfo {author}
  {\bibfnamefont {P.}~\bibnamefont {Rousseau}}, \bibinfo {author}
  {\bibfnamefont {S.}~\bibnamefont {Reed}}, \bibinfo {author} {\bibfnamefont
  {V.}~\bibnamefont {Yanovsky}}, \bibinfo {author} {\bibfnamefont {D.~W.}\
  \bibnamefont {Litzenberg}}, \bibinfo {author} {\bibfnamefont
  {K.}~\bibnamefont {Krushelnick}}, \ and\ \bibinfo {author} {\bibfnamefont
  {A.}~\bibnamefont {Maksimchuk}},\ }\href {\doibase
  10.1103/PhysRevE.78.026412} {\bibfield  {journal} {\bibinfo  {journal} {Phys.
  Rev. E}\ }\textbf {\bibinfo {volume} {78}},\ \bibinfo {pages} {026412}
  (\bibinfo {year} {2008})}\BibitemShut {NoStop}%
\bibitem [{\citenamefont {Chen}\ \emph {et~al.}(2008)\citenamefont {Chen},
  \citenamefont {Pukhov}, \citenamefont {Sheng},\ and\ \citenamefont
  {Yan}}]{MChenLaser}%
  \BibitemOpen
  \bibfield  {author} {\bibinfo {author} {\bibfnamefont {M.}~\bibnamefont
  {Chen}}, \bibinfo {author} {\bibfnamefont {A.}~\bibnamefont {Pukhov}},
  \bibinfo {author} {\bibfnamefont {Z.~M.}\ \bibnamefont {Sheng}}, \ and\
  \bibinfo {author} {\bibfnamefont {X.~Q.}\ \bibnamefont {Yan}},\ }\href
  {\doibase 10.1063/1.3019105} {\bibfield  {journal} {\bibinfo  {journal}
  {Phys. Plasmas}\ }\textbf {\bibinfo {volume} {15}},\ \bibinfo {pages}
  {113103} (\bibinfo {year} {2008})}\BibitemShut {NoStop}%
\bibitem [{\citenamefont {Chen}\ \emph {et~al.}(2009)\citenamefont {Chen},
  \citenamefont {Pukhov}, \citenamefont {Yu},\ and\ \citenamefont
  {Sheng}}]{MChenShaped}%
  \BibitemOpen
  \bibfield  {author} {\bibinfo {author} {\bibfnamefont {M.}~\bibnamefont
  {Chen}}, \bibinfo {author} {\bibfnamefont {A.}~\bibnamefont {Pukhov}},
  \bibinfo {author} {\bibfnamefont {T.~P.}\ \bibnamefont {Yu}}, \ and\ \bibinfo
  {author} {\bibfnamefont {Z.~M.}\ \bibnamefont {Sheng}},\ }\href {\doibase
  10.1103/PhysRevLett.103.024801} {\bibfield  {journal} {\bibinfo  {journal}
  {Phys. Rev. Lett.}\ }\textbf {\bibinfo {volume} {103}},\ \bibinfo {pages}
  {024801} (\bibinfo {year} {2009})}\BibitemShut {NoStop}%
\bibitem [{\citenamefont {Chen}\ \emph {et~al.}(2011)\citenamefont {Chen},
  \citenamefont {Kumar}, \citenamefont {Pukhov},\ and\ \citenamefont
  {Yu}}]{MChenStabilised}%
  \BibitemOpen
  \bibfield  {author} {\bibinfo {author} {\bibfnamefont {M.}~\bibnamefont
  {Chen}}, \bibinfo {author} {\bibfnamefont {N.}~\bibnamefont {Kumar}},
  \bibinfo {author} {\bibfnamefont {A.}~\bibnamefont {Pukhov}}, \ and\ \bibinfo
  {author} {\bibfnamefont {T.-P.}\ \bibnamefont {Yu}},\ }\href {\doibase
  10.1063/1.3606562} {\bibfield  {journal} {\bibinfo  {journal} {Phys.
  Plasmas}\ }\textbf {\bibinfo {volume} {18}},\ \bibinfo {pages} {073106}
  (\bibinfo {year} {2011})}\BibitemShut {NoStop}%
\bibitem [{\citenamefont {Yu}\ \emph {et~al.}(2010)\citenamefont {Yu},
  \citenamefont {Pukhov}, \citenamefont {Shvets},\ and\ \citenamefont
  {Chen}}]{Tyu}%
  \BibitemOpen
  \bibfield  {author} {\bibinfo {author} {\bibfnamefont {T.-P.}\ \bibnamefont
  {Yu}}, \bibinfo {author} {\bibfnamefont {A.}~\bibnamefont {Pukhov}}, \bibinfo
  {author} {\bibfnamefont {G.}~\bibnamefont {Shvets}}, \ and\ \bibinfo {author}
  {\bibfnamefont {M.}~\bibnamefont {Chen}},\ }\href {\doibase
  10.1103/PhysRevLett.105.065002} {\bibfield  {journal} {\bibinfo  {journal}
  {Phys. Rev. Lett.}\ }\textbf {\bibinfo {volume} {105}},\ \bibinfo {pages}
  {065002} (\bibinfo {year} {2010})}\BibitemShut {NoStop}%
\bibitem [{\citenamefont {Shen}\ \emph {et~al.}(2017)\citenamefont {Shen},
  \citenamefont {Qiao}, \citenamefont {Zhang}, \citenamefont {Kar},
  \citenamefont {Zhou}, \citenamefont {Chang}, \citenamefont {Borghesi},\ and\
  \citenamefont {He}}]{XFShen}%
  \BibitemOpen
  \bibfield  {author} {\bibinfo {author} {\bibfnamefont {X.~F.}\ \bibnamefont
  {Shen}}, \bibinfo {author} {\bibfnamefont {B.}~\bibnamefont {Qiao}}, \bibinfo
  {author} {\bibfnamefont {H.}~\bibnamefont {Zhang}}, \bibinfo {author}
  {\bibfnamefont {S.}~\bibnamefont {Kar}}, \bibinfo {author} {\bibfnamefont
  {C.~T.}\ \bibnamefont {Zhou}}, \bibinfo {author} {\bibfnamefont {H.~X.}\
  \bibnamefont {Chang}}, \bibinfo {author} {\bibfnamefont {M.}~\bibnamefont
  {Borghesi}}, \ and\ \bibinfo {author} {\bibfnamefont {X.~T.}\ \bibnamefont
  {He}},\ }\href {\doibase 10.1103/PhysRevLett.118.204802} {\bibfield
  {journal} {\bibinfo  {journal} {Phys. Rev. Lett.}\ }\textbf {\bibinfo
  {volume} {118}},\ \bibinfo {pages} {204802} (\bibinfo {year}
  {2017})}\BibitemShut {NoStop}%
\bibitem [{\citenamefont {Macchi}\ \emph {et~al.}(2005)\citenamefont {Macchi},
  \citenamefont {Cattani}, \citenamefont {Liseykina},\ and\ \citenamefont
  {Cornolti}}]{AMacchiIonBunches}%
  \BibitemOpen
  \bibfield  {author} {\bibinfo {author} {\bibfnamefont {A.}~\bibnamefont
  {Macchi}}, \bibinfo {author} {\bibfnamefont {F.}~\bibnamefont {Cattani}},
  \bibinfo {author} {\bibfnamefont {T.~V.}\ \bibnamefont {Liseykina}}, \ and\
  \bibinfo {author} {\bibfnamefont {F.}~\bibnamefont {Cornolti}},\ }\href
  {\doibase 10.1103/PhysRevLett.94.165003} {\bibfield  {journal} {\bibinfo
  {journal} {Phys. Rev. Lett.}\ }\textbf {\bibinfo {volume} {94}},\ \bibinfo
  {pages} {165003} (\bibinfo {year} {2005})}\BibitemShut {NoStop}%
\bibitem [{\citenamefont {Robinson}\ \emph {et~al.}(2008)\citenamefont
  {Robinson}, \citenamefont {Zepf}, \citenamefont {Kar}, \citenamefont
  {Evans},\ and\ \citenamefont {Bellei}}]{APLRnjp}%
  \BibitemOpen
  \bibfield  {author} {\bibinfo {author} {\bibfnamefont {A.~P.~L.}\
  \bibnamefont {Robinson}}, \bibinfo {author} {\bibfnamefont {M.}~\bibnamefont
  {Zepf}}, \bibinfo {author} {\bibfnamefont {S.}~\bibnamefont {Kar}}, \bibinfo
  {author} {\bibfnamefont {R.~G.}\ \bibnamefont {Evans}}, \ and\ \bibinfo
  {author} {\bibfnamefont {C.}~\bibnamefont {Bellei}},\ }\href
  {http://stacks.iop.org/1367-2630/10/i=1/a=013021} {\bibfield  {journal}
  {\bibinfo  {journal} {New J. Phys.}\ }\textbf {\bibinfo {volume} {10}},\
  \bibinfo {pages} {013021} (\bibinfo {year} {2008})}\BibitemShut {NoStop}%
\bibitem [{\citenamefont {Klimo}\ \emph {et~al.}(2008)\citenamefont {Klimo},
  \citenamefont {Psikal}, \citenamefont {Limpouch},\ and\ \citenamefont
  {Tikhonchuk}}]{OKlimo}%
  \BibitemOpen
  \bibfield  {author} {\bibinfo {author} {\bibfnamefont {O.}~\bibnamefont
  {Klimo}}, \bibinfo {author} {\bibfnamefont {J.}~\bibnamefont {Psikal}},
  \bibinfo {author} {\bibfnamefont {J.}~\bibnamefont {Limpouch}}, \ and\
  \bibinfo {author} {\bibfnamefont {V.~T.}\ \bibnamefont {Tikhonchuk}},\ }\href
  {\doibase 10.1103/PhysRevSTAB.11.031301} {\bibfield  {journal} {\bibinfo
  {journal} {Phys. Rev. ST Accel. Beams}\ }\textbf {\bibinfo {volume} {11}},\
  \bibinfo {pages} {031301} (\bibinfo {year} {2008})}\BibitemShut {NoStop}%
\bibitem [{\citenamefont {Robinson}\ \emph {et~al.}(2009)\citenamefont
  {Robinson}, \citenamefont {Gibbon}, \citenamefont {Zepf}, \citenamefont
  {Kar}, \citenamefont {Evans},\ and\ \citenamefont {Bellei}}]{APLRppcfHB}%
  \BibitemOpen
  \bibfield  {author} {\bibinfo {author} {\bibfnamefont {A.~P.~L.}\
  \bibnamefont {Robinson}}, \bibinfo {author} {\bibfnamefont {P.}~\bibnamefont
  {Gibbon}}, \bibinfo {author} {\bibfnamefont {M.}~\bibnamefont {Zepf}},
  \bibinfo {author} {\bibfnamefont {S.}~\bibnamefont {Kar}}, \bibinfo {author}
  {\bibfnamefont {R.~G.}\ \bibnamefont {Evans}}, \ and\ \bibinfo {author}
  {\bibfnamefont {C.}~\bibnamefont {Bellei}},\ }\href {\doibase
  10.1088/0741-3335/51/2/024004} {\bibfield  {journal} {\bibinfo  {journal}
  {Plasma Physics and Controlled Fusion}\ }\textbf {\bibinfo {volume} {51}},\
  \bibinfo {pages} {024004} (\bibinfo {year} {2009})}\BibitemShut {NoStop}%
\bibitem [{\citenamefont {Liseykina}\ \emph {et~al.}(2008)\citenamefont
  {Liseykina}, \citenamefont {Borghesi}, \citenamefont {Macchi},\ and\
  \citenamefont {Tuveri}}]{LiseykinaCP}%
  \BibitemOpen
  \bibfield  {author} {\bibinfo {author} {\bibfnamefont {T.~V.}\ \bibnamefont
  {Liseykina}}, \bibinfo {author} {\bibfnamefont {M.}~\bibnamefont {Borghesi}},
  \bibinfo {author} {\bibfnamefont {A.}~\bibnamefont {Macchi}}, \ and\ \bibinfo
  {author} {\bibfnamefont {S.}~\bibnamefont {Tuveri}},\ }\href
  {http://stacks.iop.org/0741-3335/50/124033} {\bibfield  {journal} {\bibinfo
  {journal} {Plasma Phys. Control. Fusion}\ }\textbf {\bibinfo {volume} {50}},\
  \bibinfo {pages} {124033} (\bibinfo {year} {2008})}\BibitemShut {NoStop}%
\bibitem [{\citenamefont {Tamburini}\ \emph {et~al.}(2010)\citenamefont
  {Tamburini}, \citenamefont {Pegoraro}, \citenamefont {Piazza}, \citenamefont
  {Keitel},\ and\ \citenamefont {Macchi}}]{Tamburini_2010}%
  \BibitemOpen
  \bibfield  {author} {\bibinfo {author} {\bibfnamefont {M.}~\bibnamefont
  {Tamburini}}, \bibinfo {author} {\bibfnamefont {F.}~\bibnamefont {Pegoraro}},
  \bibinfo {author} {\bibfnamefont {A.~D.}\ \bibnamefont {Piazza}}, \bibinfo
  {author} {\bibfnamefont {C.~H.}\ \bibnamefont {Keitel}}, \ and\ \bibinfo
  {author} {\bibfnamefont {A.}~\bibnamefont {Macchi}},\ }\href {\doibase
  10.1088/1367-2630/12/12/123005} {\bibfield  {journal} {\bibinfo  {journal}
  {New Journal of Physics}\ }\textbf {\bibinfo {volume} {12}},\ \bibinfo
  {pages} {123005} (\bibinfo {year} {2010})}\BibitemShut {NoStop}%
\bibitem [{\citenamefont {Esirkepov}\ \emph {et~al.}(2004)\citenamefont
  {Esirkepov}, \citenamefont {Borghesi}, \citenamefont {Bulanov}, \citenamefont
  {Mourou},\ and\ \citenamefont {Tajima}}]{Esirkepov}%
  \BibitemOpen
  \bibfield  {author} {\bibinfo {author} {\bibfnamefont {T.}~\bibnamefont
  {Esirkepov}}, \bibinfo {author} {\bibfnamefont {M.}~\bibnamefont {Borghesi}},
  \bibinfo {author} {\bibfnamefont {S.~V.}\ \bibnamefont {Bulanov}}, \bibinfo
  {author} {\bibfnamefont {G.}~\bibnamefont {Mourou}}, \ and\ \bibinfo {author}
  {\bibfnamefont {T.}~\bibnamefont {Tajima}},\ }\href {\doibase
  10.1103/PhysRevLett.92.175003} {\bibfield  {journal} {\bibinfo  {journal}
  {Phys. Rev. Lett.}\ }\textbf {\bibinfo {volume} {92}},\ \bibinfo {pages}
  {175003} (\bibinfo {year} {2004})}\BibitemShut {NoStop}%
\bibitem [{Note1()}]{Note1}%
  \BibitemOpen
  \bibinfo {note} {Note that in the ultrarelativistic case the relation $a(\phi
  ) < \zeta '(\phi )$ allows to increase the laser intensity or decrease the
  foil areal density maintaining $\protect \mathcal {R} \approx 1$, which, in
  principle, allows to attain ion energies well beyond those attainable with
  the bound $a_0 \lesssim \zeta $.}\BibitemShut {Stop}%
\bibitem [{Note2()}]{Note2}%
  \BibitemOpen
  \bibinfo {note} {The key condition for this approximation to hold is $n_e \gg
  a_0 n_c$ and $\ell \ll \lambda $.}\BibitemShut {Stop}%
\bibitem [{\citenamefont {Derouillat}\ \emph {et~al.}(2018)\citenamefont
  {Derouillat}, \citenamefont {Beck}, \citenamefont {Pérez}, \citenamefont
  {Vinci}, \citenamefont {Chiaramello}, \citenamefont {Grassi}, \citenamefont
  {Flé}, \citenamefont {Bouchard}, \citenamefont {Plotnikov}, \citenamefont
  {Aunai}, \citenamefont {Dargent}, \citenamefont {Riconda},\ and\
  \citenamefont {Grech}}]{SmileiPaper}%
  \BibitemOpen
  \bibfield  {author} {\bibinfo {author} {\bibfnamefont {J.}~\bibnamefont
  {Derouillat}}, \bibinfo {author} {\bibfnamefont {A.}~\bibnamefont {Beck}},
  \bibinfo {author} {\bibfnamefont {F.}~\bibnamefont {Pérez}}, \bibinfo
  {author} {\bibfnamefont {T.}~\bibnamefont {Vinci}}, \bibinfo {author}
  {\bibfnamefont {M.}~\bibnamefont {Chiaramello}}, \bibinfo {author}
  {\bibfnamefont {A.}~\bibnamefont {Grassi}}, \bibinfo {author} {\bibfnamefont
  {M.}~\bibnamefont {Flé}}, \bibinfo {author} {\bibfnamefont {G.}~\bibnamefont
  {Bouchard}}, \bibinfo {author} {\bibfnamefont {I.}~\bibnamefont {Plotnikov}},
  \bibinfo {author} {\bibfnamefont {N.}~\bibnamefont {Aunai}}, \bibinfo
  {author} {\bibfnamefont {J.}~\bibnamefont {Dargent}}, \bibinfo {author}
  {\bibfnamefont {C.}~\bibnamefont {Riconda}}, \ and\ \bibinfo {author}
  {\bibfnamefont {M.}~\bibnamefont {Grech}},\ }\href {\doibase
  https://doi.org/10.1016/j.cpc.2017.09.024} {\bibfield  {journal} {\bibinfo
  {journal} {Computer Physics Communications}\ }\textbf {\bibinfo {volume}
  {222}},\ \bibinfo {pages} {351 } (\bibinfo {year} {2018})}\BibitemShut
  {NoStop}%
\bibitem [{\citenamefont {Grech}\ \emph {et~al.}(2011)\citenamefont {Grech},
  \citenamefont {Skupin}, \citenamefont {Diaw}, \citenamefont {Schlegel},\ and\
  \citenamefont {Tikhonchuk}}]{MGrechRPA}%
  \BibitemOpen
  \bibfield  {author} {\bibinfo {author} {\bibfnamefont {M.}~\bibnamefont
  {Grech}}, \bibinfo {author} {\bibfnamefont {S.}~\bibnamefont {Skupin}},
  \bibinfo {author} {\bibfnamefont {A.}~\bibnamefont {Diaw}}, \bibinfo {author}
  {\bibfnamefont {T.}~\bibnamefont {Schlegel}}, \ and\ \bibinfo {author}
  {\bibfnamefont {V.~T.}\ \bibnamefont {Tikhonchuk}},\ }\href
  {http://stacks.iop.org/1367-2630/13/i=12/a=123003} {\bibfield  {journal}
  {\bibinfo  {journal} {New J. Phys.}\ }\textbf {\bibinfo {volume} {13}},\
  \bibinfo {pages} {123003} (\bibinfo {year} {2011})}\BibitemShut {NoStop}%
\bibitem [{\citenamefont {Markey}\ \emph {et~al.}(2010)\citenamefont {Markey},
  \citenamefont {McKenna}, \citenamefont {Brenner}, \citenamefont {Carroll},
  \citenamefont {G\"unther}, \citenamefont {Harres}, \citenamefont {Kar},
  \citenamefont {Lancaster}, \citenamefont {N\"urnberg}, \citenamefont {Quinn},
  \citenamefont {Robinson}, \citenamefont {Roth}, \citenamefont {Zepf},\ and\
  \citenamefont {Neely}}]{KMarkey}%
  \BibitemOpen
  \bibfield  {author} {\bibinfo {author} {\bibfnamefont {K.}~\bibnamefont
  {Markey}}, \bibinfo {author} {\bibfnamefont {P.}~\bibnamefont {McKenna}},
  \bibinfo {author} {\bibfnamefont {C.~M.}\ \bibnamefont {Brenner}}, \bibinfo
  {author} {\bibfnamefont {D.~C.}\ \bibnamefont {Carroll}}, \bibinfo {author}
  {\bibfnamefont {M.~M.}\ \bibnamefont {G\"unther}}, \bibinfo {author}
  {\bibfnamefont {K.}~\bibnamefont {Harres}}, \bibinfo {author} {\bibfnamefont
  {S.}~\bibnamefont {Kar}}, \bibinfo {author} {\bibfnamefont {K.}~\bibnamefont
  {Lancaster}}, \bibinfo {author} {\bibfnamefont {F.}~\bibnamefont
  {N\"urnberg}}, \bibinfo {author} {\bibfnamefont {M.~N.}\ \bibnamefont
  {Quinn}}, \bibinfo {author} {\bibfnamefont {A.~P.~L.}\ \bibnamefont
  {Robinson}}, \bibinfo {author} {\bibfnamefont {M.}~\bibnamefont {Roth}},
  \bibinfo {author} {\bibfnamefont {M.}~\bibnamefont {Zepf}}, \ and\ \bibinfo
  {author} {\bibfnamefont {D.}~\bibnamefont {Neely}},\ }\href {\doibase
  10.1103/PhysRevLett.105.195008} {\bibfield  {journal} {\bibinfo  {journal}
  {Phys. Rev. Lett.}\ }\textbf {\bibinfo {volume} {105}},\ \bibinfo {pages}
  {195008} (\bibinfo {year} {2010})}\BibitemShut {NoStop}%
\bibitem [{\citenamefont {Naumova}\ \emph {et~al.}(2009)\citenamefont
  {Naumova}, \citenamefont {Schlegel}, \citenamefont {Tikhonchuk},
  \citenamefont {Labaune}, \citenamefont {Sokolov},\ and\ \citenamefont
  {Mourou}}]{PhysRevLett.102.025002}%
  \BibitemOpen
  \bibfield  {author} {\bibinfo {author} {\bibfnamefont {N.}~\bibnamefont
  {Naumova}}, \bibinfo {author} {\bibfnamefont {T.}~\bibnamefont {Schlegel}},
  \bibinfo {author} {\bibfnamefont {V.~T.}\ \bibnamefont {Tikhonchuk}},
  \bibinfo {author} {\bibfnamefont {C.}~\bibnamefont {Labaune}}, \bibinfo
  {author} {\bibfnamefont {I.~V.}\ \bibnamefont {Sokolov}}, \ and\ \bibinfo
  {author} {\bibfnamefont {G.}~\bibnamefont {Mourou}},\ }\href {\doibase
  10.1103/PhysRevLett.102.025002} {\bibfield  {journal} {\bibinfo  {journal}
  {Phys. Rev. Lett.}\ }\textbf {\bibinfo {volume} {102}},\ \bibinfo {pages}
  {025002} (\bibinfo {year} {2009})}\BibitemShut {NoStop}%
\bibitem [{\citenamefont {Sinigardi}\ \emph {et~al.}(2013)\citenamefont
  {Sinigardi}, \citenamefont {Turchetti}, \citenamefont {Londrillo},
  \citenamefont {Rossi}, \citenamefont {Giove}, \citenamefont {De~Martinis},\
  and\ \citenamefont {Sumini}}]{SSinigardi}%
  \BibitemOpen
  \bibfield  {author} {\bibinfo {author} {\bibfnamefont {S.}~\bibnamefont
  {Sinigardi}}, \bibinfo {author} {\bibfnamefont {G.}~\bibnamefont
  {Turchetti}}, \bibinfo {author} {\bibfnamefont {P.}~\bibnamefont
  {Londrillo}}, \bibinfo {author} {\bibfnamefont {F.}~\bibnamefont {Rossi}},
  \bibinfo {author} {\bibfnamefont {D.}~\bibnamefont {Giove}}, \bibinfo
  {author} {\bibfnamefont {C.}~\bibnamefont {De~Martinis}}, \ and\ \bibinfo
  {author} {\bibfnamefont {M.}~\bibnamefont {Sumini}},\ }\href {\doibase
  10.1103/PhysRevSTAB.16.031301} {\bibfield  {journal} {\bibinfo  {journal}
  {Phys. Rev. ST Accel. Beams}\ }\textbf {\bibinfo {volume} {16}},\ \bibinfo
  {pages} {031301} (\bibinfo {year} {2013})}\BibitemShut {NoStop}%
\end{thebibliography}%

\end{document}